\documentclass[letterpaper, 10 pt, conference]{ieeeconf}  
\usepackage{relsize}
\usepackage{multicol}
\usepackage{subcaption}
\usepackage{graphicx}
\usepackage{amssymb}
\usepackage{flushend}

\usepackage{mathrsfs}
 \usepackage{mathtools, nccmath}

\usepackage{setspace}
\usepackage{amsmath,mathtools}
\usepackage{widetext}
\usepackage[usenames,dvipsnames]{color} 
\usepackage{ifthen}

\usepackage{tabularx}
\usepackage{longtable}
\usepackage{listings}
\usepackage{booktabs}
\usepackage{float}
\usepackage{makecell}
\usepackage{epstopdf}
\usepackage{pifont} 
\usepackage{wasysym}
\usepackage{tikz}
\usepackage{upgreek}
\usepackage[normalem]{ulem}
\usepackage{flafter}  
 
\usepackage{amsthm}
\usepackage[top=55pt,bottom=55pt,left=55pt,right=55pt]{geometry}
\newtheorem{thm}{Theorem}

\newtheorem{remark}[thm]{Remark}
\theoremstyle{plain}
\usepackage{lmodern,pdftexcmds,amsmath}
\usepackage[section,above,below]{placeins}
\usepackage{ifthen}

\newboolean{showcomments}
\setboolean{showcomments}{true}

\newcommand{\dennice}[1]{\ifthenelse{\boolean{showcomments}}
{\textcolor{cyan}{Dennice says: #1}}{}}
\newcommand{\addcites}[0]{\ifthenelse{\boolean{showcomments}}
{\textcolor{magenta}{(add cite(s))}}{}}
\newcommand{\chang}[1]{\ifthenelse{\boolean{showcomments}}
{\textcolor{red}{Chang says: #1}}{}}
\DeclareMathAlphabet{\mathsfbi}{OT1}{\sfdefault}{bx}{sl}
\DeclareMathVersion{sfletters}
\SetSymbolFont{letters}{sfletters}{OML}{ntxsfmi}{b}{it}

\newtheorem{definition}[thm]{Definition} 

\makeatletter
\newcommand{\mathbfsbilow}[1]{%
  \text{\mathversion{sfletters}$\m@th#1$}%
}
\makeatother

\IEEEoverridecommandlockouts                              

\title{\LARGE \bf
A structured input-output approach to characterizing optimal perturbations in wall-bounded shear flows}

\author{Chang Liu$^{1}$, Yu Shuai$^{2}$, Aishwarya Rath$^{3}$, and Dennice F. Gayme$^{3}$
\thanks{$^{1}$ Department of Physics,
        University of California, Berkeley, CA 94720, USA
        {\tt\small chang\_liu@berkeley.edu}}%
\thanks{$^{2}$ Department of Mechanical and Aerospace Engineering, Princeton University,
        Princeton, NJ 45435, USA
        {\tt\small yu\_shuai@princeton.edu}}%
        
\thanks{$^{3}$ Department of Mechanical Engineering, Johns Hopkins University,
        Baltimore, MD 21218, USA
        {\tt\small \{arath9,dennice\}@jhu.edu}}%
}

\begin{document}

\maketitle
\thispagestyle{empty}
\pagestyle{empty}

\begin{abstract}

This work builds upon recent work exploiting the notion of structured singular values to capture nonlinear interactions in the analysis of wall-bounded shear flows. In this context, the structured uncertainty can be interpreted in terms of the flow structures most likely to be amplified (the optimal perturbations). Here we further analyze these perturbations through a problem reformulation that decomposes this uncertainty into three components associated with the streamwise, wall-normal and spanwise velocity correlations. We then demonstrate that the structural features of these correlations are consistent with nonlinear optimal perturbations and results from secondary stability analysis associated with streamwise streaks. These results indicate the potential of structured input-output analysis for gaining insight into both linear and nonlinear behavior that can be used to inform sensing and control strategies for transitional wall-bounded shear flows.

\end{abstract}
	
\section{Introduction}

Input-output analysis (e.g., $\mathcal{H}_2$ and $\mathcal{H}_\infty$ analysis of the spatio-temporal frequency response operator associated with the linearized Navier Stokes equations) has been widely employed to characterize energy amplification and the dominant structural features in both transitional and turbulent wall-bounded shear flows, see e.g. the review articles \cite{schmid2007nonmodal,mckeon2017engine,jovanovic2020bypass,zare2020stochastic}. A substantial benefit of these approaches in studying transitional flows versus eigenvalue analysis  is their ability  to capture transient growth mechanisms that have been shown to play a role in subcritical transition \cite{trefethen1993hydrodynamic,farrell1993,jovanovic2001spatio,jovanovic2003frequency}. For example, these methods have been used to highlight the importance of streamwise elongated structures, as well as to identify the spacing of the streamwise vortices and streaks, which play an important role  in the dynamics of the flow, see e.g., \cite{Bamieh2001,farrell1993,del2006linear,hwang2010linear}. 
Jovanovi\'c and Bamieh \cite{Jovanovic2005} used a decomposition of the input-output pairs mapping body forcing applied to each of the three velocity components to the componentwise energy density that allowed them to characterize the importance of cross-stream forcing in redistributing background mean shear across the channel height \cite{Jovanovic2005}. The associated analysis of streamwise streak formation provided new insight into the `lift-up mechanism' that was already known to be important in energy growth and organization of the flow \cite{ellingsen1975stability,Brandt2014,jovanovic2020bypass}. However, comparisons with  studies based on direct numerical simulations (DNS) of the full nonlinear Navier-Stokes equations~\cite{reddy1998stability}, experiments \cite{prigent2003long}, and nonlinear methods \cite{Rabin2012,farano2015hairpin} have indicated that linear analysis tends to over-emphasize this lift-up effect~\cite{duguet2013minimal,Brandt2014} and thereby obscure the contributions of streamwise varying structures.  

In order to provide a more complete characterization of the flow, a number of researchers have sought to include nonlinear effects in the input-output approach; e.g., through harmonic balance methods \cite{rigas2021nonlinear} that build upon techniques for analyzing systems with spatio-temporal periodic coefficients \cite{jovanovic2008h2,fardad2008frequency,padovan2020analysis,padovan2022analysis,bamieh1991lifting}. Nonlinearity has also been included in stability analysis using quadratic constraints within a linear matrix inequality formulations \cite{ahmadi2019framework,liu2020input,kalur2021estimating,kalur2021nonlinear,toso2022regional}.  Liu \& Gayme~\cite{liu2021structuredJournal} proposed an alternative approach that employs an input-output model of the nonlinearity  placed within a feedback interconnection with the linearized dynamics (in the spirit of a Lur\'e decomposition \cite{kalman1963lyapunov,khalil2002nonlinear} of the problem \cite{sharma2006stabilising,mckeon2017engine}). They then reformulate the spatio-temporal response operator to isolate the unknown input-output gain associated with the model of the nonlinearity as a structured singular value \cite{doyle1982analysis,packard1993complex,zhou1996robust}.  They refer to this approach as structured input-output analysis (SIOA), as the model of the nonlinearity restricts the `structured uncertainty' (gain operator) to be block diagonal to mirror the form of the nonlinearity in the Navier-Stokes equations. SIOA has been shown to provide the  required weakening of the lift-up mechanism \cite[\S 3.3]{liu2021structuredJournal} that enables the analysis to recover the streamwise varying structures shown most likely to trigger transition in experiments, DNS and nonlinear optimal perturbation (NLOP) analysis \cite{prigent2003long,Rabin2012,reddy1998stability,farano2015hairpin}. It has also been used to identify both the horizontal length scales and inclination angles associated with the oblique laminar-turbulent patterns observed in transitioning channel flow \cite{liu2021structuredJournal,shuai2022structured},  stably stratified plane Couette flows \cite{liu2022structured} and spanwise rotating plane Couette flow \cite{liu2021feedback}. SIOA has also shown promise in reproducing characteristic time scales in channel flows~\cite{shuai2022structured}, as well as in extracting dominant forcing and response modes \cite{mushtaq2023structured} that can be used to inform control strategies.  
 
This work further develops the SIOA approach to provide additional physical insight regarding the block diagonal `structured uncertainty' operator, which describes the perturbations most likely to destabilize the feedback interconnection in the input-output sense \cite{zhou1996robust}. In order to interrogate this operator, we first  reformulate the feedback interconnection originally proposed in \cite{liu2021structuredJournal} to  isolate the three velocity components associated with the structured uncertainty operator. We interpret the resulting full-block components as velocity correlations between two wall-normal locations associated with the largest structured response; a number of studies e.g., \cite{Zare2017,liu2020input,towne2020resolvent,nogueira2021forcing} have shown the benefits of analyzing similar correlations in different contexts.  We then study the component-wise velocity profiles and the wall-normal variation of these structures.
We refer to these velocity fields as
``optimal perturbations''  in analogy with NLOP and  linear analysis aimed at identifying perturbations leading to the largest energy growth \cite{butler1993optimal}. 

The results demonstrate that the velocity correlations and the profile of the autocorrelation for plane Couette flow show a maximum magnitude near the channel center, which is consistent with the behavior of the eigenfunction associated with the secondary instability of streamwise streaks \cite{reddy1998stability}.  The real parts of these autocorrelations also show the same sign reversal near the channel center as the NLOP \cite{cherubini2013nonlinear}. In plane Poiseuille flow, the destabilizing velocity autocorrelation instead vanishes near the channel center, which is consistent with both the behavior of the NLOP \cite{parente2022linear} and results analyzing optimal secondary energy growth  \cite{cossu2007optimal}. In contrast, linear optimal perturbations of plane Poiseuille flow peak near the channel center; see e.g. the comparison in \cite{parente2022linear}. The agreement of these results with those from various nonlinear analysis approaches provides further evidence that behavior associated with nonlinear effects can be captured using SIOA-based approaches.

In what follows, \S \ref{sec:SSV_loop} describes the feedback interconnection used to compute the components of the ``optimal perturbations''. The structured response and associated structured uncertainty are analyzed in \S\ref{sec:Structured_response} and \S\ref{sec:SSV_covariance_matrix}, respectively. We then conclude the paper and discuss future work in \S\ref{sec:conclusion}. 

\section{Structured input-output analysis}
\label{sec:SSV_loop}
In this section, we describe the  spatio-temporal frequency response operator of the linearized Navier-Stokes equations and formulate the model of the nonlinearity. We then place them in a feedback interconnection structure  that enables us to analyze the perturbations that are most likely to induce transition using the structured singular value formalism \cite{packard1993complex,zhou1996robust}.

We consider two types of flow between two infinite horizontal parallel plates; plane Couette flow, which is driven by the relative motion of the plates, and plane Poiseuille flow, which is pressure driven. The coordinates $x$, $y$ and $z$ respectively define the streamwise, wall-normal and spanwise directions. The three components of the velocity vector field  $\boldsymbol{u}_T(x,y,z,t)=\begin{bmatrix} u_T & v_T & w_T \end{bmatrix}^{\text{T}}$ are respectively associated with the $x$, $y$ and $z$ directions. 

For each flow configuration, we decompose the velocity field $\boldsymbol{u}_T$ into a laminar base flow $(U, W, W)$, where $V=W=0$ and $U(y)=y$ for plane Couette flow and $U(y)=1-y^2$ for plane Poiseuille flow, and fluctuations $\boldsymbol{u}=(u,v,w)$ about the base flow.  The dynamics of the velocity fluctuations $\boldsymbol{u}$  are governed by the non-dimensional Navier-Stokes equations:
\begin{align}
\label{eq:NS_All}
\partial_{t} \boldsymbol{u}  
+  U\partial_x \boldsymbol{u}  + v\,U'\boldsymbol{e}_x +\boldsymbol{\nabla} p
-\frac{1}{Re}{\nabla}^2 \boldsymbol{u}
 &=- \boldsymbol{u} \!\cdot\! \boldsymbol{\nabla} \boldsymbol{u}, 
\end{align}
and $\boldsymbol{\nabla} \!\cdot\! \boldsymbol{u}=0$. Here, $U':=dU(y)/dy$ and $p(x,y,z,t)$ denotes the pressure fluctuations associated with the decomposed pressure field $p_T=P+p$. The Reynolds number is $Re=U_nh/\nu$, where $\nu$ is the kinematic viscosity and $h$ is channel half-height. Here, $\pm U_n$ is the velocity at the channel walls for plane Couette flow and $U_n$ is the channel centerline velocity for plane Poiseuille flow. We impose no-slip and no penetration boundary conditions; i.e., $\boldsymbol{u}(y=\pm 1)=\boldsymbol{0}$.

In order to build the feedback interconnection of interest, we model the nonlinear terms (nonlinearity) $-\boldsymbol{u}\!\cdot\! \boldsymbol{\nabla }\boldsymbol{u}$ in \eqref{eq:NS_All} as a forcing:
\begin{subequations}
     \label{eq:f_uncertain_model}
\begin{align}
\boldsymbol{f}_\xi:=-\boldsymbol{u}_\xi \!\cdot\! \boldsymbol{\nabla }\boldsymbol{u}
=\begin{bmatrix}-\boldsymbol{u}_\xi \! \cdot\! \boldsymbol{\nabla} u\\ -\boldsymbol{u}_\xi\!\cdot\! \boldsymbol{\nabla} v\\ -\boldsymbol{u}_\xi\!\cdot\! \boldsymbol{\nabla} w\end{bmatrix}\!=:\!\begin{bmatrix}f_{x,\xi}\\
f_{y,\xi}\\
f_{z,\xi}\end{bmatrix}.  
\end{align}
\end{subequations}
\noindent In this input-output model of the nonlinearity, $-\boldsymbol{u}_{\xi}=-[u_\xi,\,v_\xi,\,w_\xi]^\text{T}$ acts as a gain mapping the gradient of each velocity component to the corresponding component of the modeled forcing driving the linearized dynamics. 

We use the standard practice of  transforming the forced linearized dynamics to a divergence-free reference frame, where the new states are the wall-normal velocity $v$ and wall-normal vorticity $\omega_y:=\partial_z u-\partial_x w$, see details in e.g., \cite{Bamieh2001}.  We then exploit the shift-invariance of these flows in the $(x,z,t)$ directions to perform a triple Fourier transform of $v$, $\omega_y$ and the components of the forcing model $f_{x,\xi}$, $f_{y,\xi}$, and $f_{z,\xi}$, where for a variable $\psi$,  
 $\widehat{\psi}(y;k_x,k_z,\omega):=\int\limits_{-\infty}^{\infty} \int\limits_{-\infty}^{\infty}\int\limits_{-\infty}^{\infty}\psi(x,y,z,t)e^{-\text{i}(k_x x + k_z z+\omega t )}\,dx\,dz\,dt$. Here $\text{i}=\sqrt{-1}$ is the imaginary unit,  $\omega$ is the temporal frequency, and  $k_x$ and $k_z$ are the respective dimensionless $x$ and $z$ wavenumbers. 

\begin{figure*}[t]

	(a) \hspace{0.7\textwidth} (b)
	
    \centering
\includegraphics[width=0.65\textwidth]{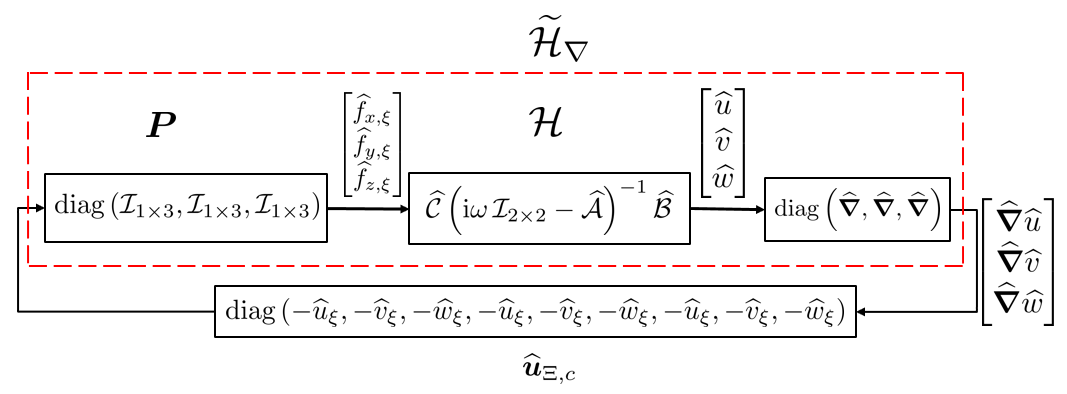}
    \hspace{0.2in}\includegraphics[width=0.8in,trim=-0 -0.5in 0 0]{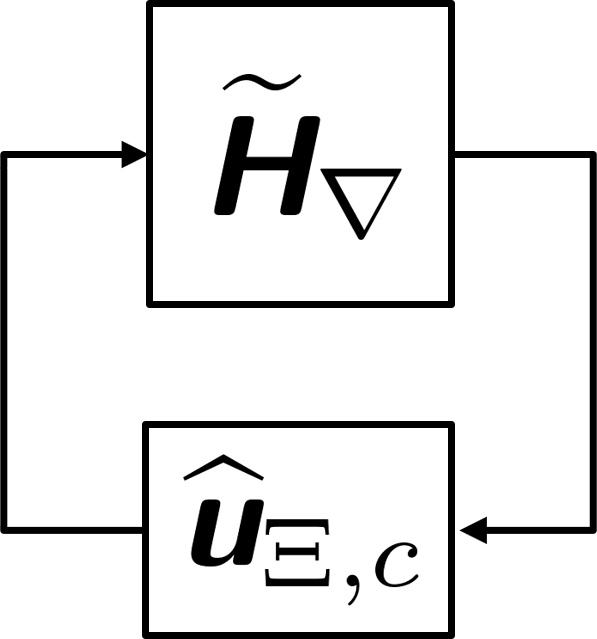}
    \caption{  (a) Block diagram describing the SIOA feedback interconnection structure. Panel (b) redraws panel (a) after discretization.}
    \label{fig:feedback_detail}
\end{figure*}
The resulting equations describing the transformed linearized equations subject to the modeled forcing  are
\begin{subequations}
\label{eq:ABC_frequency}
\begin{equation}
    \text{i}\omega\begin{bmatrix}
    \widehat{v}\\
    \widehat{\omega}_y
    \end{bmatrix}=\widehat{\mathcal{A}}\begin{bmatrix}
    \widehat{v}\\
    \widehat{\omega}_y
    \end{bmatrix}+\widehat{\mathcal{B}}\begin{bmatrix}\widehat{f}_{x,\xi}\\
    \widehat{f}_{y,\xi}\\
    \widehat{f}_{z,\xi}\end{bmatrix},\;\;
    \begin{bmatrix}\widehat{u}\\
    \widehat{v}\\
    \widehat{w}\end{bmatrix}=\widehat{\mathcal{C}}\begin{bmatrix}
    \widehat{v}\\
    \widehat{\omega}_y
    \end{bmatrix}.\tag{\theequation a,b}
\end{equation}
\end{subequations}
The operators in \eqref{eq:ABC_frequency} are defined in the standard way, see e.g. \cite{Jovanovic2005},
\begin{subequations}
\label{eq:operator_ABC}
\begin{align}
    \widehat{\mathcal{A}}:=&\widehat{\mathcal{M}}^{-1}\begin{bmatrix}
    \mathcal{L}_{11} & 0\\
    -\text{i}k_zU' & \mathcal{L}_{22}
    \end{bmatrix},\;\widehat{\mathcal{M}}:=\begin{bmatrix} \widehat{\nabla}^2 & 0\\   0 & \mathcal{I} \end{bmatrix},\label{eq:operator_ABC_A} \\
    \mathcal{\widehat{B}}:=&\widehat{\mathcal{M}}^{-1}
    \begin{bmatrix}
    -\text{i}k_x\partial_y & -(k_x^2+k_z^2) & -\text{i}k_z \partial_y\\
    \text{i}k_z & 0 & -\text{i}k_x
    \end{bmatrix},\label{eq:operator_ABC_B}\\
    \mathcal{\widehat{C}}:=&\frac{1}{k_x^2+k_z^2}\begin{bmatrix}
    \text{i}k_x\partial_y & -\text{i}k_z\\
    k_x^2+k_z^2 & 0\\
    \text{i}k_z \partial_y & \text{i}k_x
    \end{bmatrix},\label{eq:operator_ABC_C}
\end{align}
\end{subequations}
where $\mathcal{L}_{11}:=-\text{i}k_xU{\widehat{\nabla}}^2+\text{i}k_xU''+\widehat{{\nabla}}^4/Re$ and $\mathcal{L}_{22}:=-\text{i}k_x U+\widehat{{\nabla}}^2/Re$. The associated boundary conditions are $\widehat{v}(y=\pm 1)=
    \frac{\partial \widehat{v}}{\partial y}(y=\pm1)=\widehat{\omega}_y(y=\pm 1)=0$. 
    
    We write the transformed forcing model as
\begin{align}
\begin{bmatrix}
    \widehat{f}_{x,\xi}\\
    \widehat{f}_{y,\xi}\\
    \widehat{f}_{z,\xi}
    \end{bmatrix}=&\boldsymbol{P}\; \boldsymbol{\widehat{u}}_{\Xi,c}\text{diag}\left(\boldsymbol{\widehat{\nabla}},\boldsymbol{\widehat{\nabla}},\boldsymbol{\widehat{\nabla}}\right)\begin{bmatrix}
\widehat{u}\\
\widehat{v}\\
\widehat{w}
\end{bmatrix},\;\;\text{where}
\label{eq:feedback_structured_uncertainty}\\
\boldsymbol{P}:=&\text{diag}\left(\mathcal{I}_{1\times 3},\mathcal{I}_{1\times 3},\mathcal{I}_{1\times 3}\right),\label{eq:P}\\ \boldsymbol{\widehat{u}}_{\Xi,c}:=&\mathcal{I}_{3\times 3}\otimes\text{diag}\left(-\widehat{u}_\xi,-\widehat{v}_\xi,-\widehat{w}_\xi \right),
    \label{eq:uncertainty_u_Xi_c}
\end{align}
and $\otimes$ denotes Kronecker product. Here, $\mathcal{I}_{1\times 3}:=[\mathcal{I},\mathcal{I},\mathcal{I}]$ and $\mathcal{I}_{3\times 3}:=\text{diag}(\mathcal{I},\mathcal{I},\mathcal{I})$, where $\mathcal{I}$ is the identity operator and $\text{diag}(\cdot)$ indicates a block diagonal operation. This decomposition allows us to isolate a block-diagonal operator $\widehat{\boldsymbol{u}}_{\Xi,c}$, which is the structured uncertainty that we will investigate in this work.

The spatio-temporal frequency response operator of the system in \eqref{eq:ABC_frequency}, defined as \begin{equation}
    \mathcal{H}(y;k_x,k_z,\omega):=\widehat{\mathcal{C}}\left(\text{i}\omega\,\mathcal{I}_{2\times 2}-\widehat{\mathcal{A}}\right)^{-1}\widehat{\mathcal{B}}
     \label{eq:linearized_ABC}
\end{equation}
maps the input forcing to the velocity vector at the same spatial-temporal wavenumber-frequency triplet; i.e., $\boldsymbol{\widehat{u}}(y;k_x,k_z,\omega)=\mathcal{H}(y;k_x,k_z,\omega)\boldsymbol{\widehat{f}}_{\xi}(y;k_x,k_z,\omega)$.

In order to isolate the structured uncertainty $-\widehat{\boldsymbol{u}}_{\Xi,c}$ that we seek to analyze, we combine the linear gradient operator and matrix $\boldsymbol{P}$ with the spatio-temporal frequency response $\mathcal{H}$ to obtain a modified operator \begin{align}\mathcal{\widetilde{H}}_{\nabla}(y;k_x,k_z,\omega\!)\!:=\!
     \text{diag}\left(\!\boldsymbol{\widehat{\nabla}},\boldsymbol{\widehat{\nabla}},\boldsymbol{\widehat{\nabla}}\right)\!\!\mathcal{H}(y;k_x,k_z, \omega)\boldsymbol{P}.
    \label{eq:H_operator_grad}
\end{align}
Fig. \ref{fig:feedback_detail}(a) shows the feedback interconnection between $\widetilde{\mathcal{H}}_{\nabla}$ in \eqref{eq:H_operator_grad} (outlined with a red dashed line) and the structured uncertainty $\widehat{\boldsymbol{u}}_{\Xi,c}$ in \eqref{eq:uncertainty_u_Xi_c}. The slight reformulations of this feedback interconnection structure and forcing expression \eqref{eq:feedback_structured_uncertainty} versus those in \cite{liu2021structuredJournal} provide a decomposition of $\widehat{\boldsymbol{u}}_{\Xi,c}$ into the three   components $\widehat{u}_{\xi}$, $\widehat{v}_{\xi}$ and $\widehat{w}_{\xi}$, which enables analysis of characteristics associated with the three velocity components.

The operators in equation \eqref{eq:operator_ABC} are discretized using the Chebyshev collocation method with derivatives computed using the Chebyshev differential matrices generated by the MATLAB routines of \cite{Weideman2000}. The number of wall-normal grid points is denoted as $N_y$. Fig. \ref{fig:feedback_detail}(b) shows the discretized version of the feedback interconnection between the modified spatio-temporal frequency response and the structured uncertainty, where $\mathsfbi{\widetilde{H}}_{\nabla}$ and $\mathbfsbilow{\widehat{u}}_{\Xi,c}$ respectively represent the numerical discretizations in the wall-normal direction of  $\widetilde{\mathcal{H}}_{\nabla}$ in \eqref{eq:H_operator_grad} and $\boldsymbol{\widehat{u}}_{\Xi,c}$ in \eqref{eq:uncertainty_u_Xi_c}.

We characterize the perturbations associated with the most amplified flow structures under structured forcing in terms of the structured singular value associated with $\widetilde{\mathsfbi{H}}_\nabla$. This quantity is defined following e.g., \cite[definition 3.1]{packard1993complex}.

\begin{definition}
Given the wavenumber and frequency triplet $(k_x,k_z,\omega)$, the structured singular value is defined as $\mu_{\mathbfsbilow{\widetilde{U}}_{\Xi,c}}\left[\mathbfsbilow{\widetilde{H}}_{\nabla}(k_x,k_z,\omega)\right]:=$ 
\begin{align}
    \frac{1}{\text{min}\{\bar{\sigma}[\mathbfsbilow{\widetilde{u}}_{\Xi,c}]\,:\,\mathbfsbilow{\widetilde{u}}_{\Xi,c}\in \mathbfsbilow{\widetilde{U}}_{\Xi,c},\,\text{det}[\mathsfbi{I}-\mathbfsbilow{\widetilde{H}}_{\nabla}\mathbfsbilow{\widetilde{u}}_{\Xi,c}]=0\}}.
    \label{eq:mu}
\end{align}
If no $\mathbfsbilow{\widetilde{u}}_{\Xi,c}\in \mathbfsbilow{\widetilde{U}}_{\Xi,c}$ makes $\mathsfbi{I}-\widetilde{\mathbfsbilow{H}}_{\nabla}\mathbfsbilow{\widetilde{u}}_{\Xi,c}$ singular, then $\mu_{\mathbfsbilow{\widetilde{U}}_{\Xi,c}}[\mathbfsbilow{\widetilde{H}}_{\nabla}]:=0$. Here, $\bar{\sigma}[\cdot]$ is the largest singular value, $\text{det}[\cdot]$ is the determinant of the argument, and $\mathsfbi{I}$ is the identity matrix. 
\label{def:mu}
\end{definition}

The model of the nonlinearity prescribes the set that contains structured uncertainty (i.e., the $\mathbfsbilow{\widehat{u}}_{\Xi,c}\in \mathbfsbilow{\widehat{U}}_{\Xi,c}$) as:
\begin{align}
\label{eq:uncertain_set_repeated_block}
\mathbfsbilow{\widehat{U}}_{\Xi,c}:=\Big\{&\mathsfbi{I}_{3N_y}\otimes \text{diag}\big(-\mathbfsbilow{\widehat{u}}_{\xi},-\mathbfsbilow{\widehat{v}}_{\xi},-\mathbfsbilow{\widehat{w}}_{\xi}\big)\nonumber\\
:&-\mathbfsbilow{\widehat{u}}_{\xi},-\mathbfsbilow{\widehat{v}}_{\xi},-\mathbfsbilow{\widehat{w}}_{\xi}\in \mathbb{C}^{N_y\times N_y}
\Big\},
\end{align}
where $\mathsfbi{I}_{3N_y}\in \mathbb{C}^{3N_y\times 3N_y}$ is the identity matrix with the corresponding size. However, this set involves repeated complex blocks, which is a constraint that cannot be enforced  in the off-the-shelf \texttt{mussv} command in the Robust Control Toolbox \cite{balas2005robust} in MATLAB. In order to apply this toolbox, we relax the structured uncertainty containing set to
\begin{align}
\label{eq:uncertain_set}
&\mathbfsbilow{\widetilde{U}}_{\Xi,c}:=\Big\{\text{diag}\big(-\mathbfsbilow{\widehat{u}}_{\xi,1},-\mathbfsbilow{\widehat{v}}_{\xi,1},-\mathbfsbilow{\widehat{w}}_{\xi,1},\nonumber\\
&-\mathbfsbilow{\widehat{u}}_{\xi,2},-\mathbfsbilow{\widehat{v}}_{\xi,2},-\mathbfsbilow{\widehat{w}}_{\xi,2},-\mathbfsbilow{\widehat{u}}_{\xi,3},
-\mathbfsbilow{\widehat{v}}_{\xi,3},-\mathbfsbilow{\widehat{w}}_{\xi,3}\big) \nonumber\\
:&-\mathbfsbilow{\widehat{u}}_{\xi,j},-\mathbfsbilow{\widehat{v}}_{\xi,j},-\mathbfsbilow{\widehat{w}}_{\xi,j}\in \mathbb{C}^{N_y\times N_y},\;\;(j=1,2,3)
\Big\}.
\end{align}
We then compute
\begin{align}
    \|\widetilde{\mathcal{H}}_{\nabla}\|_{\mu}(k_x,k_z):=\underset{\omega \in \mathbb{R}}{\text{sup}}\;\mu_{\mathbfsbilow{\widetilde{U}}_{\Xi,c}}\left[\mathbfsbilow{\widetilde{H}}_{\nabla}(k_x,k_z,\omega)\right]
\end{align}
for each wavenumber pair $(k_x,k_z)$ using the \texttt{mussv} command.  The arguments we employ  include the state-space model of $\mathbfsbilow{\widetilde{H}}_{\nabla}$ that samples the frequency domain adaptively. We use the \texttt{`Uf'} algorithm option.

 \begin{remark}This uncertainty set in \eqref{eq:uncertain_set} corresponds to a  \texttt{BlockStructure} argument comprising nine (not necessarily repeated) full $N_y\times N_y$ complex matrices when using the \texttt{mussv} command. In other words, there is no imposed relationship between $\mathbfsbilow{\widehat{u}}_{\xi,j},\mathbfsbilow{\widehat{v}}_{\xi,j},\mathbfsbilow{\widehat{w}}_{\xi,j}$ for different values of $j=1,2,3$, and as discussed above this is the relaxation being made to enable computation using the existing MATLAB toolbox.  Previous work suggests that this relaxed problem produces results consistent with nonlinear analysis \cite{liu2021structuredJournal,liu2022structured,shuai2022structured,liu2021feedback} but there may be further insights gained by extending the numerical techniques to enforce equality of these components. Such extensions based on the original formulation in \cite{liu2021structuredJournal} are described \cite{mushtaq2022structured}. Further quantification of the effect of this relaxation is a direction for future work. 
 \end{remark}

\begin{figure}
    \centering
    (a) \hspace{0.23\textwidth} (b) 
    
    \includegraphics[width=0.235\textwidth]{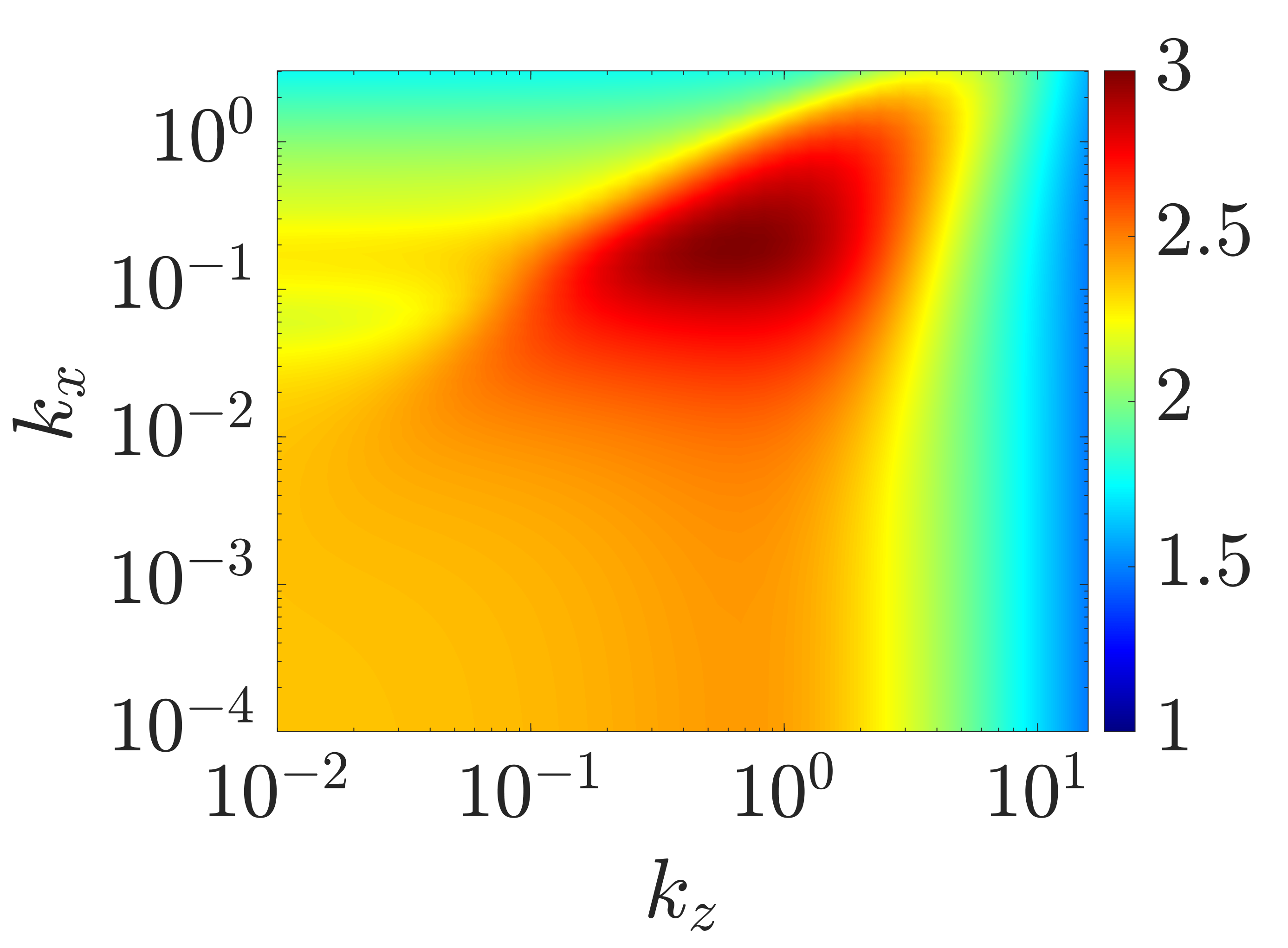}
    \includegraphics[width=0.235\textwidth]{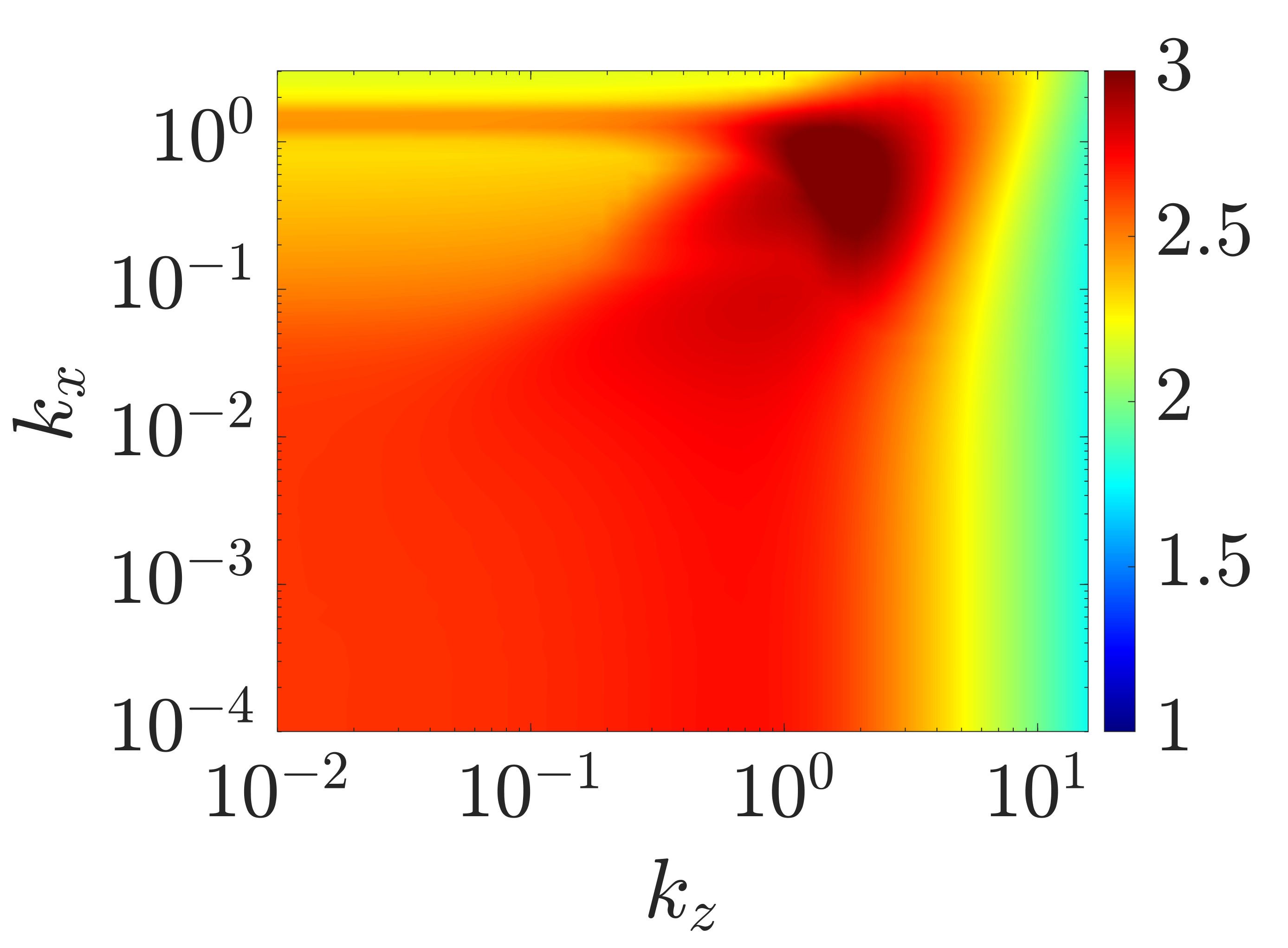}
    \caption{$\text{log}_{10}[\|\mathcal{\widetilde{H}}_{\nabla}\|_{\mu}(k_x,k_z)]$ for (a) plane Couette flow at $Re=358$ and (b) plane Poiseuille flow at $Re=690$.}
    \label{fig:mu_PCF_358}
\end{figure}

\section{Structured Response}
\label{sec:Structured_response}

Fig. \ref{fig:mu_PCF_358} shows contour plots of $\text{log}_{10}[\|\mathcal{\widetilde{H}}_{\nabla}\|_{\mu}(k_x,k_z)]$ for plane Couette flow at $Re=358$ and plane Poiseuille flow at $Re=690$, in panels (a) and (b) respectively. The results shown use 48 logarithmically  spaced grid points in $k_x\in [10^{-4}, 10^{0.48}]$,  36 logarithmically spaced grid points in $k_z\in [10^{-2}, 10^{1.2}]$ and $N_y=30$, which was shown to be adequate in the convergence study performed in \cite{liu2021structuredJournal}. As expected these results mirror those in \cite[Figs. 4(a) and 5(a)]{liu2021structuredJournal}, demonstrating that the modification to the feedback interconnection structure proposed here does not significantly affect the horizontal length scales associated with the maximal structured response. In particular,  panel (a) indicates that the largest structured response corresponds to streamwise varying flow structures consistent with the oblique waves shown to require the least energy to induce transition in DNS studies \cite{reddy1998stability}. The aspect ratios of these structures are consistent with NLOP \cite{Rabin2012} and experimental observations \cite{prigent2003long}. Panel (b) similarly identifies oblique flow structures with wavenumber pairs consistent with those identified as most likely to induce transition in DNS \cite{reddy1998stability}. The large amplitudes in the peak region and lower left quadrant of Fig. \ref{fig:mu_PCF_358}(b) are consistent with the spatially localized structures identified through NLOP analysis \cite{farano2015hairpin}. For more detail regarding these figures, see \cite[\S 3]{liu2021structuredJournal}. 

\section{Optimal Perturbation Structure}
\label{sec:SSV_covariance_matrix}

We now analyze the characteristics of optimal perturbations in terms of structured uncertainty $\widetilde{\mathbfsbilow{u}}_{\Xi,c}$. As discussed in the previous section our computation relaxes the problem such that the structured uncertainty comprises nine components $\mathbfsbilow{\widehat{u}}_{\xi,j},\mathbfsbilow{\widehat{v}}_{\xi,j},\mathbfsbilow{\widehat{w}}_{\xi,j},(j=1,2,3)$. We  are interested in the behavior of the three velocity components associated with the peak response in Fig. \ref{fig:mu_PCF_358}, so it would be preferable to have equal values for each of the three components for all $j$ allowing us to extract a single value for $\mathbfsbilow{\widehat{u}}_{\xi}$, $\mathbfsbilow{\widehat{v}}_{\xi}$ and $\mathbfsbilow{\widehat{w}}_{\xi}$. Since this is not the case, it is of interest to determine which set of the block diagonal velocity components, i.e., associated with $j=1,\,2$ or $3$, is responsible for producing the peak values in Fig. \ref{fig:mu_PCF_358}. In an effort to estimate the relative contribution of each set $j$ we 
alter the $\boldsymbol{P}$ matrix to compute the response associated with the three block diagonal triplets ($\mathbfsbilow{\widehat{u}}_{\xi,j},\mathbfsbilow{\widehat{v}}_{\xi,j},\mathbfsbilow{\widehat{w}}_{\xi,j}$) as follows:
\begin{subequations}
\begin{align}  
\widetilde{\mathcal{H}}_{\nabla,1}:=&
\widetilde{\mathcal{H}}_{\nabla}\;\text{diag}(\mathcal{I}_{3\times 3}, \boldsymbol{0}_{3\times3}, \boldsymbol{0}_{3\times 3}),\label{eq:H_nabla_1}\\
\widetilde{\mathcal{H}}_{\nabla,2}:=&
\widetilde{\mathcal{H}}_{\nabla}\;\text{diag}(\boldsymbol{0}_{3\times 3}, \mathcal{I}_{3\times3}, \boldsymbol{0}_{3\times 3}),\label{eq:H_nabla_2}\\
\widetilde{\mathcal{H}}_{\nabla,3}:=&
\widetilde{\mathcal{H}}_{\nabla}\;\text{diag}(\boldsymbol{0}_{3\times 3}, \boldsymbol{0}_{3\times3}, \mathcal{I}_{3\times 3}).\label{eq:H_nabla_3}
\end{align}
\end{subequations}
Here, the operator in \eqref{eq:H_nabla_1} corresponds to forcing only in the $x$ momentum equation, i.e., setting $\mathbfsbilow{\widehat{u}}_{\xi,j}=\boldsymbol{0},\mathbfsbilow{\widehat{v}}_{\xi,j}=\boldsymbol{0},\mathbfsbilow{\widehat{w}}_{\xi,j}=\boldsymbol{0},\;(j=2,3)$ in \eqref{eq:uncertain_set}. Similarly, \eqref{eq:H_nabla_2} corresponds to setting $\mathbfsbilow{\widehat{u}}_{\xi,j}=\boldsymbol{0},\mathbfsbilow{\widehat{v}}_{\xi,j}=\boldsymbol{0},\mathbfsbilow{\widehat{w}}_{\xi,j}=\boldsymbol{0},\;(j=1,3)$ and \eqref{eq:H_nabla_3} corresponds to $\mathbfsbilow{\widehat{u}}_{\xi,j}=\boldsymbol{0},\mathbfsbilow{\widehat{v}}_{\xi,j}=\boldsymbol{0},\mathbfsbilow{\widehat{w}}_{\xi,j}=\boldsymbol{0},\;(j=1,2)$. 

\begin{figure}
 (a) $\|\widetilde{\mathcal{H}}_{\nabla,1}\|_{\mu}$\hspace{0.05\textwidth} (b) $\|\widetilde{\mathcal{H}}_{\nabla,2}\|_{\mu}$\hspace{0.05\textwidth} (c) $\|\widetilde{\mathcal{H}}_{\nabla,3}\|_{\mu}$

    \includegraphics[width=0.155\textwidth]{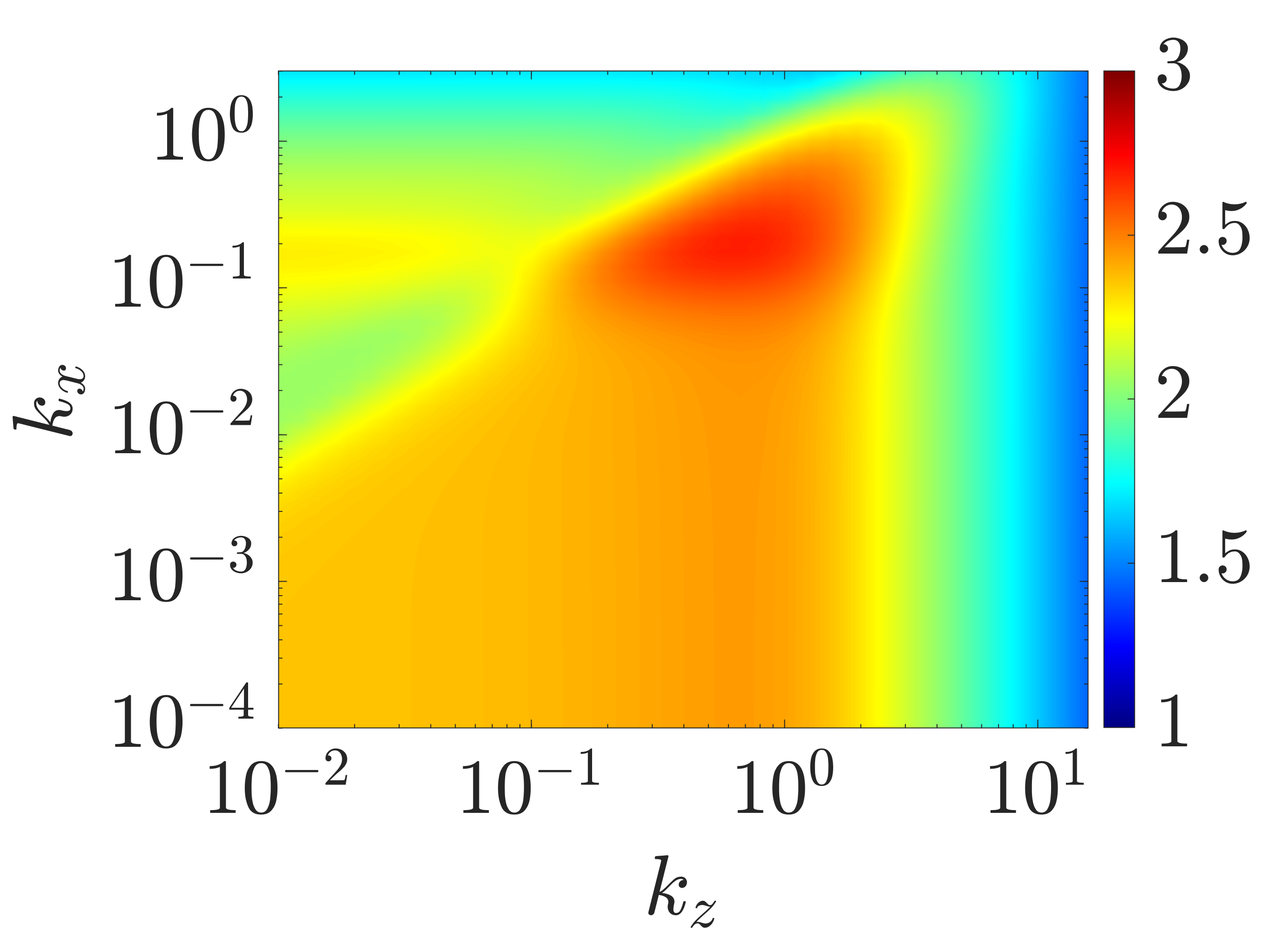}
    \includegraphics[width=0.155\textwidth]{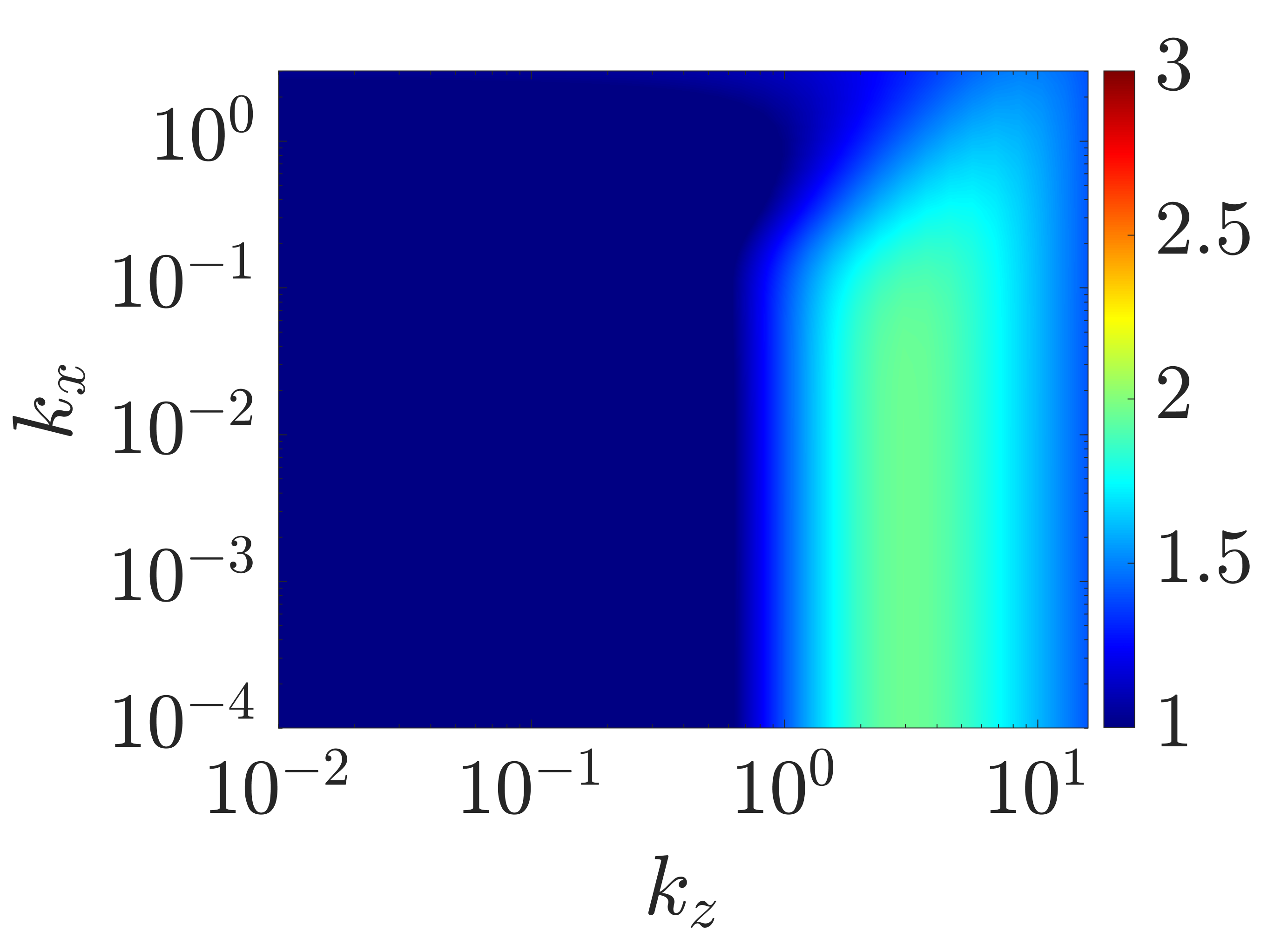}
    \includegraphics[width=0.155\textwidth]{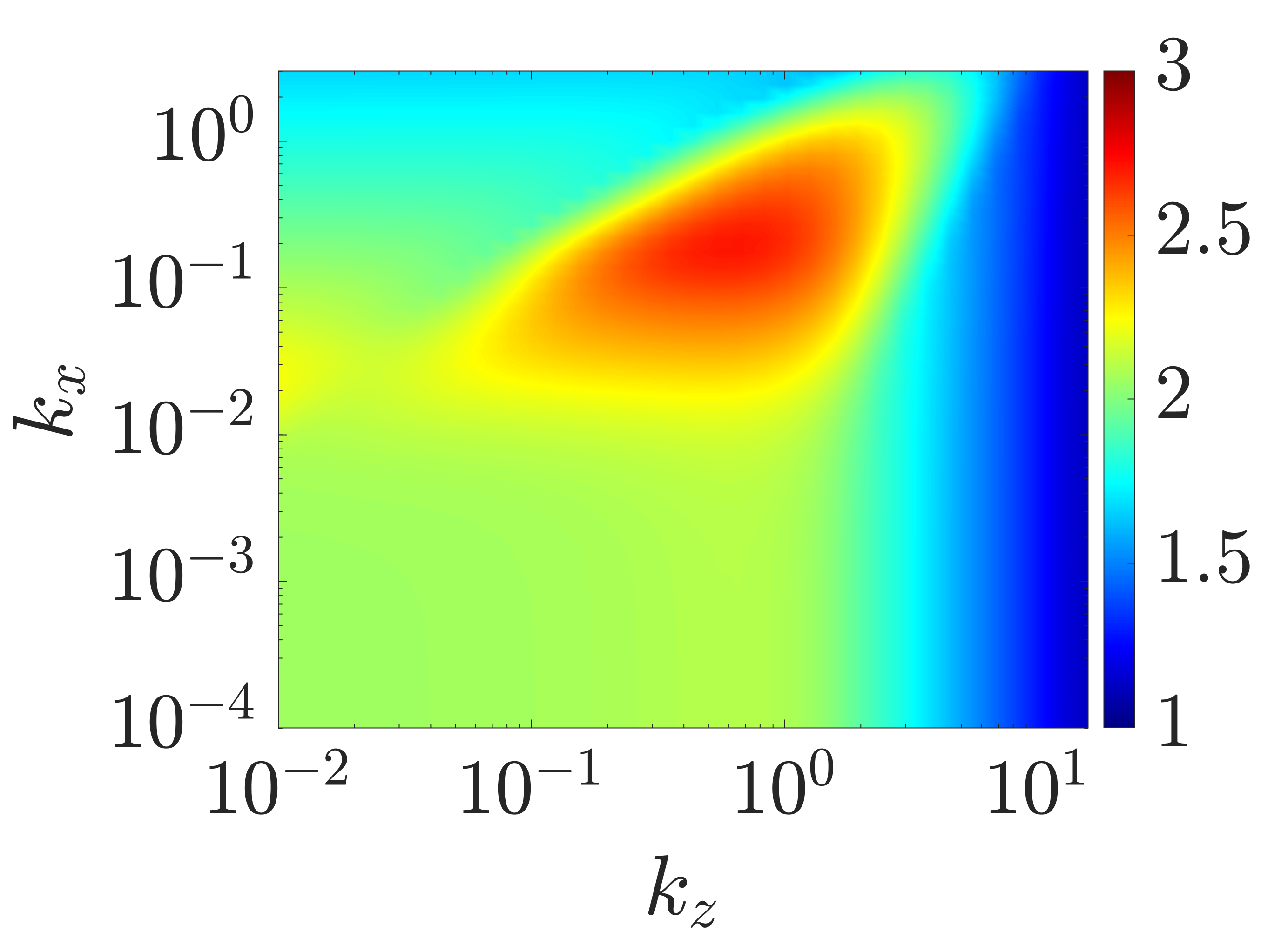}

\caption{$\text{log}_{10}[\|\widetilde{\mathcal{H}}_{\nabla,j}\|_{\mu}]$, $(j=1,2,3)$ for plane Couette flow at $Re=358$. }
    \label{fig:three_components_123_cou}
\end{figure}

\begin{figure}
 (a) $\|\widetilde{\mathcal{H}}_{\nabla,1}\|_{\mu}$\hspace{0.05\textwidth} (b) $\|\widetilde{\mathcal{H}}_{\nabla,2}\|_{\mu}$\hspace{0.05\textwidth} (c) $\|\widetilde{\mathcal{H}}_{\nabla,3}\|_{\mu}$

    \includegraphics[width=0.155\textwidth]{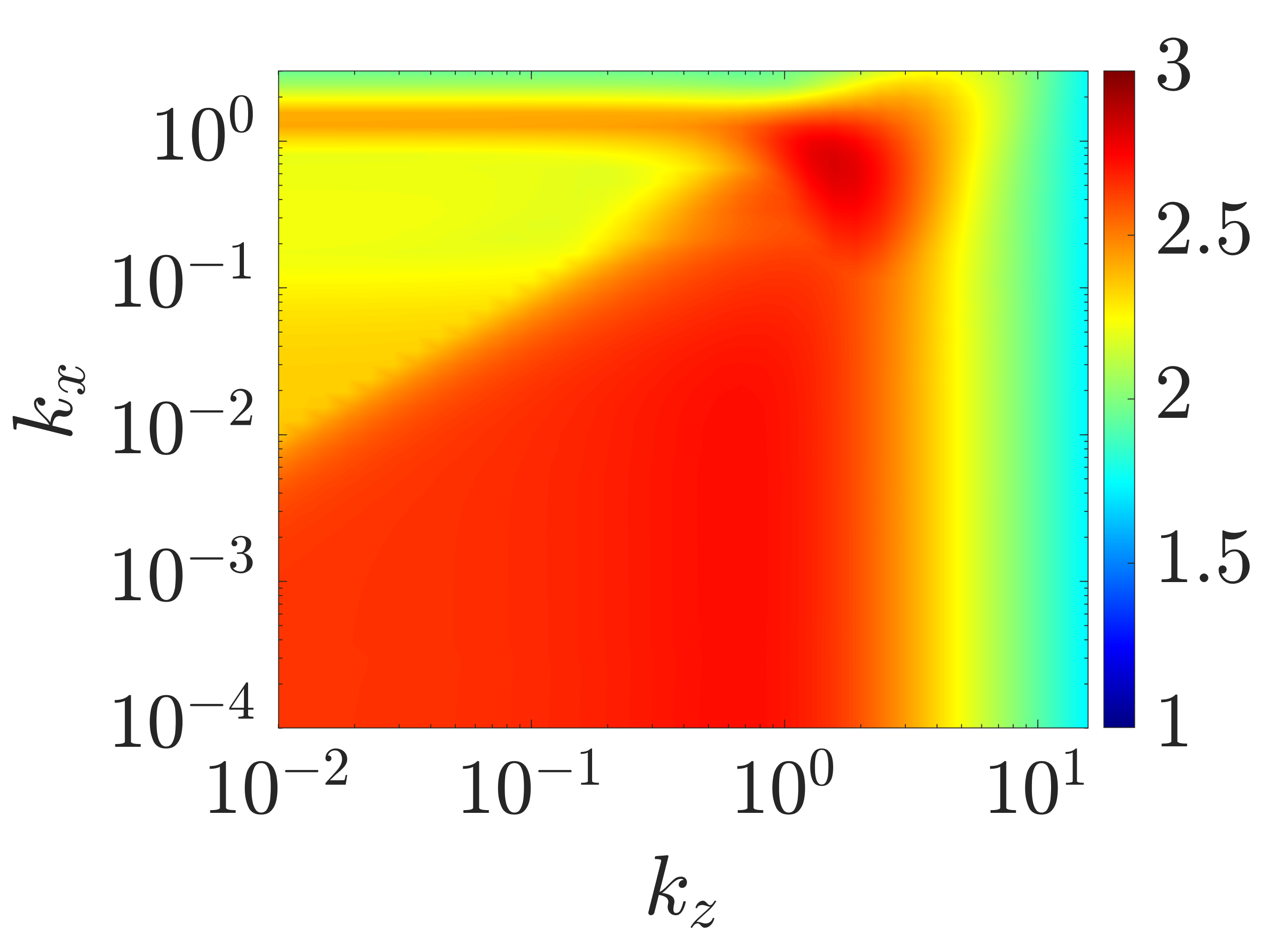}
    \includegraphics[width=0.155\textwidth]{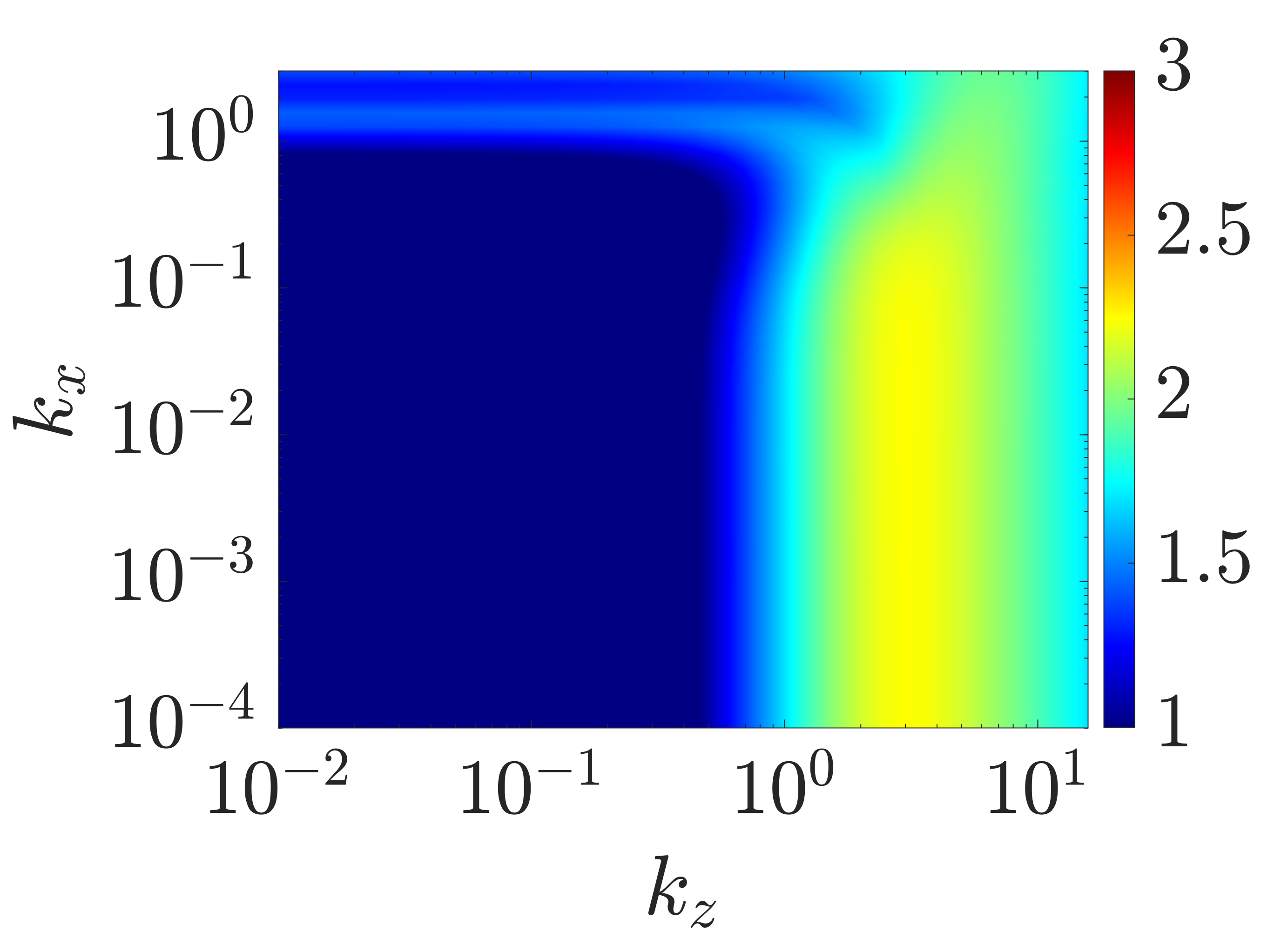}
    \includegraphics[width=0.155\textwidth]{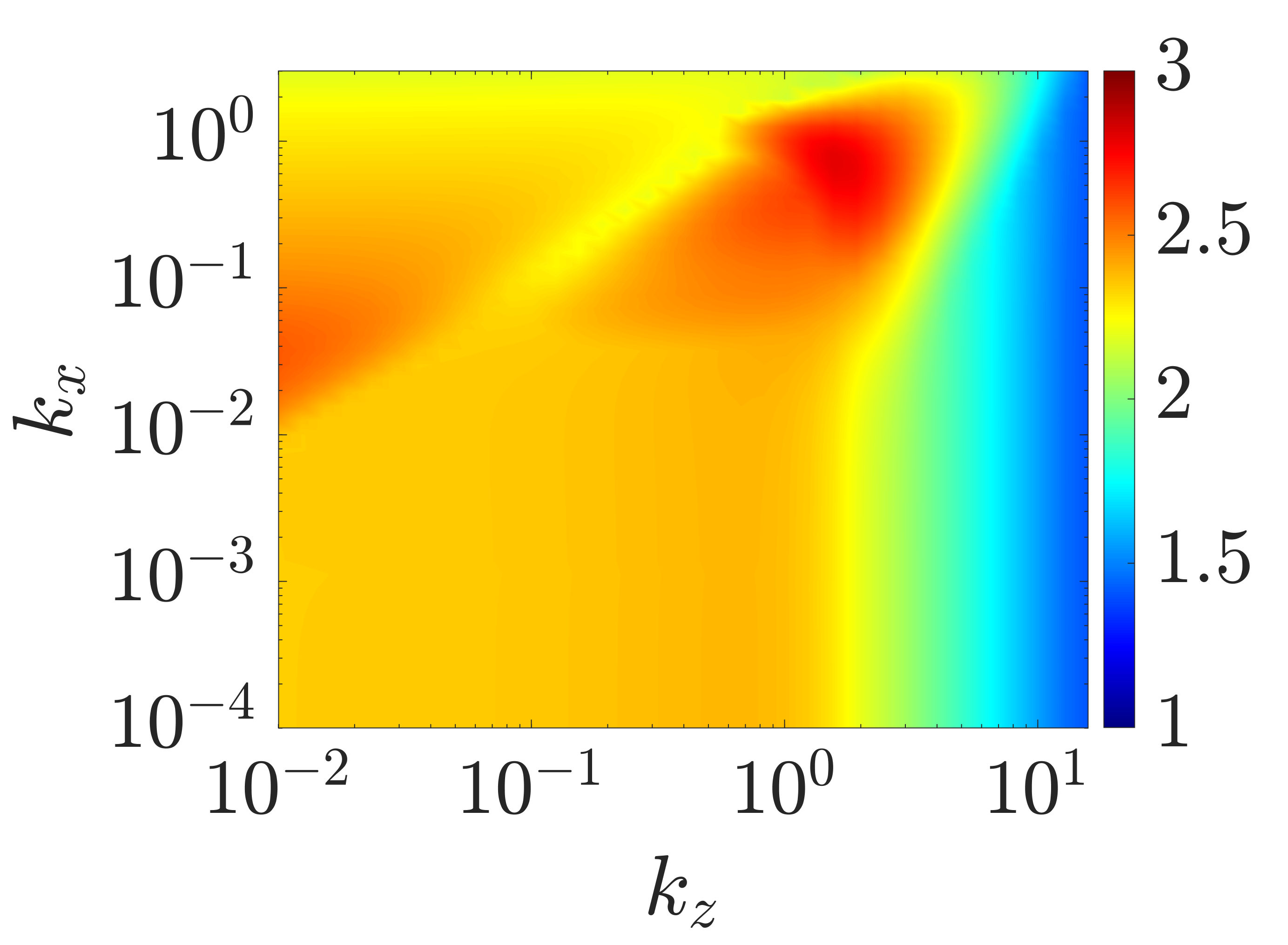}

    \caption{$\text{log}_{10}[\|\widetilde{\mathcal{H}}_{\nabla,j}\|_{\mu}]$, $(j=1,2,3)$ for plane Poiseuille flow at $Re=690$. }
    \label{fig:three_components_123_poi}
\end{figure}

Fig. \ref{fig:three_components_123_cou} shows $\|\widetilde{\mathcal{H}}_{\nabla, j}\|_{\mu}$ $(j=1,2,3)$ for plane Couette flow. These results suggest that $\|\widetilde{\mathcal{H}}_{\nabla, 1}\|_{\mu}$ and $\|\widetilde{\mathcal{H}}_{\nabla, 3}\|_{\mu}$ show a similar peak region and overall behavior to $\|\widetilde{\mathcal{H}}_{\nabla}\|_{\mu}$ in Fig. \ref{fig:mu_PCF_358}(a). Both $\|\widetilde{\mathcal{H}}_{\nabla, 1}\|_{\mu}$ and $\|\widetilde{\mathcal{H}}_{\nabla, 3}\|_{\mu}$ achieve peak values at the same $k_x=0.22$ and $k_z=0.67$, as $\|\widetilde{\mathcal{H}}_{\nabla}\|_{\mu}$ in Fig. \ref{fig:mu_PCF_358}(a). This analysis suggests that the maximum structured response is associated with the $j=1$ and $j=3$, with the response associated with $j=1$ having the most similarity to the overall response. A full analysis of how this relates to the input-output pathways in the equations is a topic of ongoing work. 

Fig. \ref{fig:three_components_123_poi} shows 
$\|\widetilde{\mathcal{H}}_{\nabla, j}\|_{\mu}$ $(j=1,2,3)$ for plane 
Poiseuille
flow. Comparing these results to those in Fig. \ref{fig:mu_PCF_358}(b) indicates similar trends to those seen for plane Couette flow, where the response associated with the first set in Fig. \ref{fig:three_components_123_poi}(a)  most closely resembles that in  Fig. \ref{fig:mu_PCF_358}(b). This response also has a peak value at the same $k_x=0.65$, $k_z=1.56$  as $\|\widetilde{\mathcal{H}}_{\nabla}\|_{\mu}$ in Fig. \ref{fig:mu_PCF_358}(b).  In this case while the peak region of $\|\widetilde{\mathcal{H}}_{\nabla, 3}\|_{\mu}$ in Fig. \ref{fig:three_components_123_poi}(c) also resembles that of  $\|\widetilde{\mathcal{H}}_{\nabla}\|_{\mu}$, it achieves the peak value at $k_x=0.81$ and $k_z=1.56$. This analysis suggests that the maximal structured response is likely associated with the set $\mathbfsbilow{\widehat{u}}_{\xi,1},\mathbfsbilow{\widehat{v}}_{\xi,1},\mathbfsbilow{\widehat{w}}_{\xi,1}$. 

We compute the matrix $\widetilde{\mathbfsbilow{u}}_{\Xi,c}(y,y';k_x, k_z, \omega)\in \mathbfsbilow{\widetilde{U}}_{\Xi,c}$ based on  \eqref{eq:H_operator_grad} and
 \eqref{eq:uncertain_set} using \texttt{mussvextract} in the Robust Control Toolbox \cite{balas2005robust}, which outputs this value as \texttt{VDelta}. Our computations use $N_y=120$ and all values in this section are multiplied by $10^3$ for ease of visualization. The Clenshaw--Curtis quadrature \cite[chapter 12]{trefethen2000spectral} is  implemented on the frequency response operator and structured uncertainty to ensure that the resulting $\widetilde{\mathbfsbilow{u}}_{\Xi,c}$ is independent of the number of Chebyshev spaced wall-normal grid points. The nine components of this block diagonal matrix $\mathbfsbilow{\widetilde{u}}_{\Xi,c}$ are full blocks corresponding to an input-output mapping at two different wall-normal locations, which we denote as $y$ and $y'$. These blocks can therefore be interpreted as spatial velocity correlations in the wall-normal direction.  Based on the results in Figs. \ref{fig:three_components_123_cou} and \ref{fig:three_components_123_poi}, we focus on the three elements $\mathbfsbilow{\widehat{u}}_{\xi,1},\mathbfsbilow{\widehat{v}}_{\xi,1},\mathbfsbilow{\widehat{w}}_{\xi,1}$, although all nine components were computed. In all of the results in this section, we use $|\cdot|$ to denote absolute value and employ $\mathcal{R}e[\cdot]$ and $\mathcal{I}m[\cdot]$ to indicate the respective real and imaginary parts of the complex entries of $\widetilde{\mathbfsbilow{u}}_{\Xi,c}(y,y';k_x, k_z, \omega)$.

\begin{figure}
    \centering
    
    (a) $\mathcal{R}e[10^3\,\widehat{\mathbfsbilow{u}}_{\xi,1}(y,y')]$ \hspace{0.09\textwidth} (b) $\mathcal{I}m[10^3\,\widehat{\mathbfsbilow{u}}_{\xi,1}(y,y')]$
    
    \includegraphics[width=0.23\textwidth]{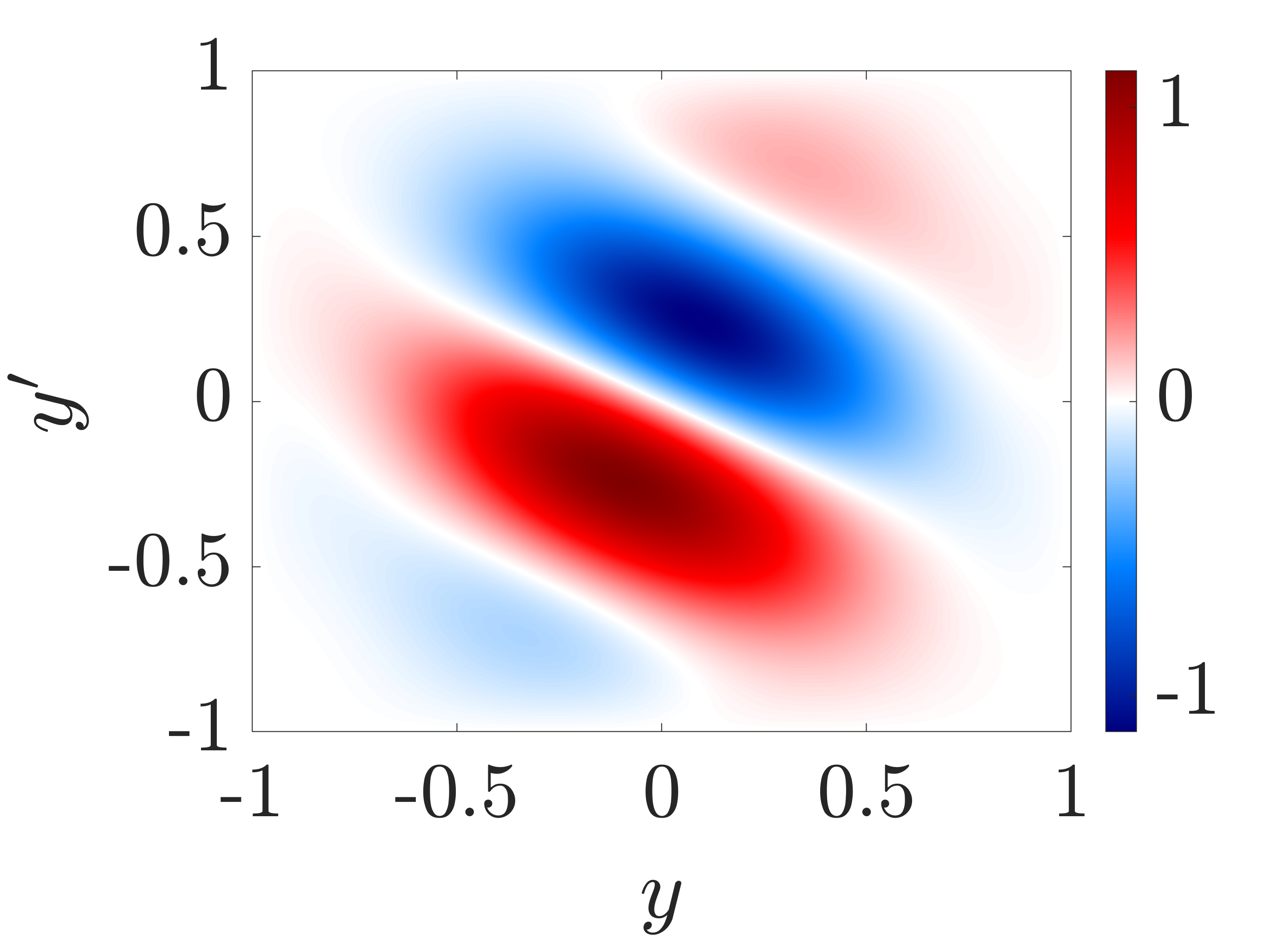}
    \includegraphics[width=0.23\textwidth]{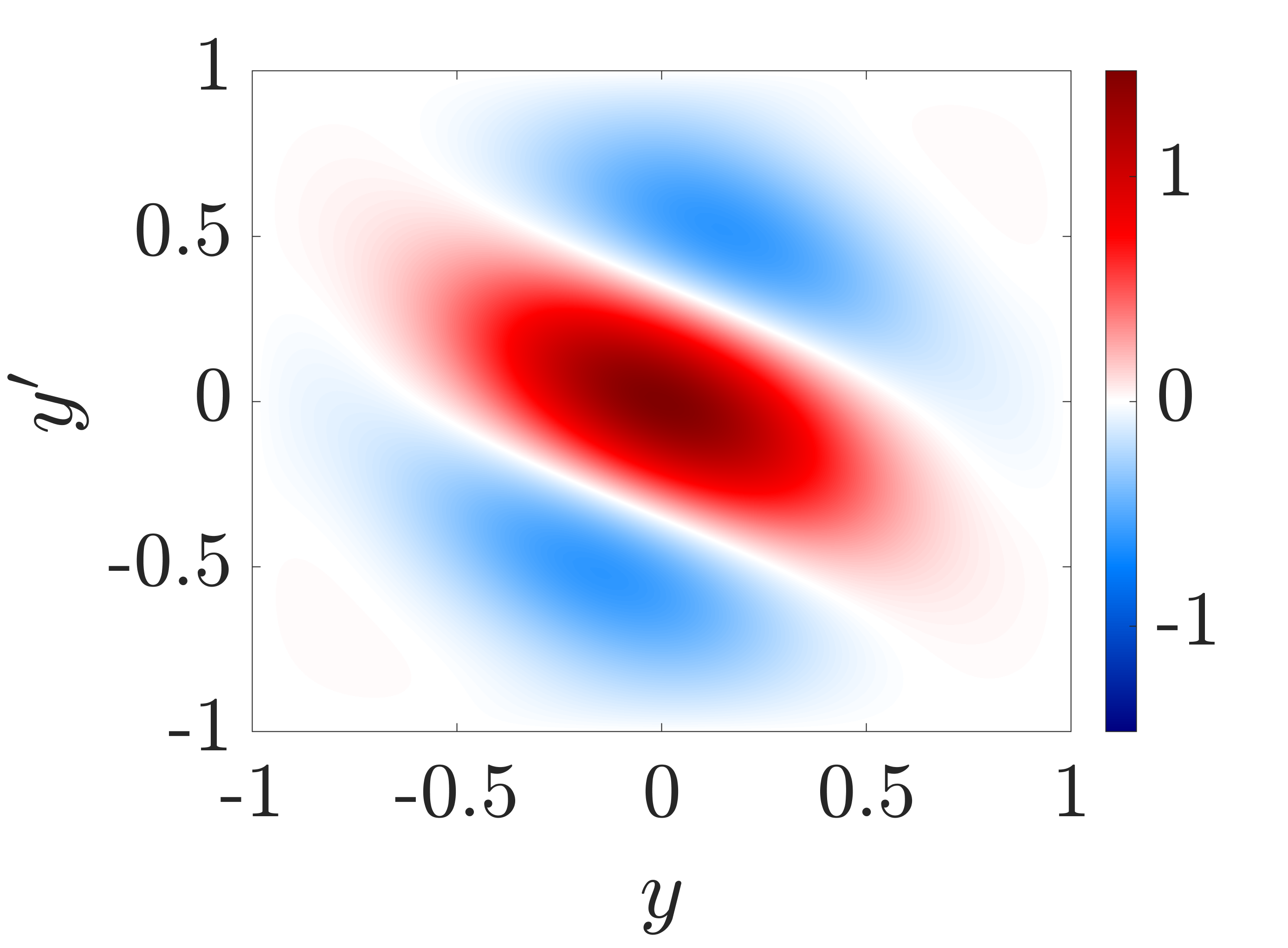}    \caption{$10^3\,\widehat{\mathbfsbilow{u}}_{\xi,1}$ for plane Couette flow at $Re=358$, $k_x=0.22$, $k_z=0.67$ and $\omega=0$.}
    \label{fig:u_xi_abs_real_imag_diag_1_cou}
\end{figure}

\begin{figure}
    \centering
    (a) $10^3\,\widehat{\mathbfsbilow{u}}_{\xi,1}$\hspace{0.05\textwidth} (b) $10^3\,\widehat{\mathbfsbilow{v}}_{\xi,1}$ \hspace{0.05\textwidth} (c) $10^3\,\widehat{\mathbfsbilow{w}}_{\xi,1}$
    
    \includegraphics[width=0.135\textwidth]{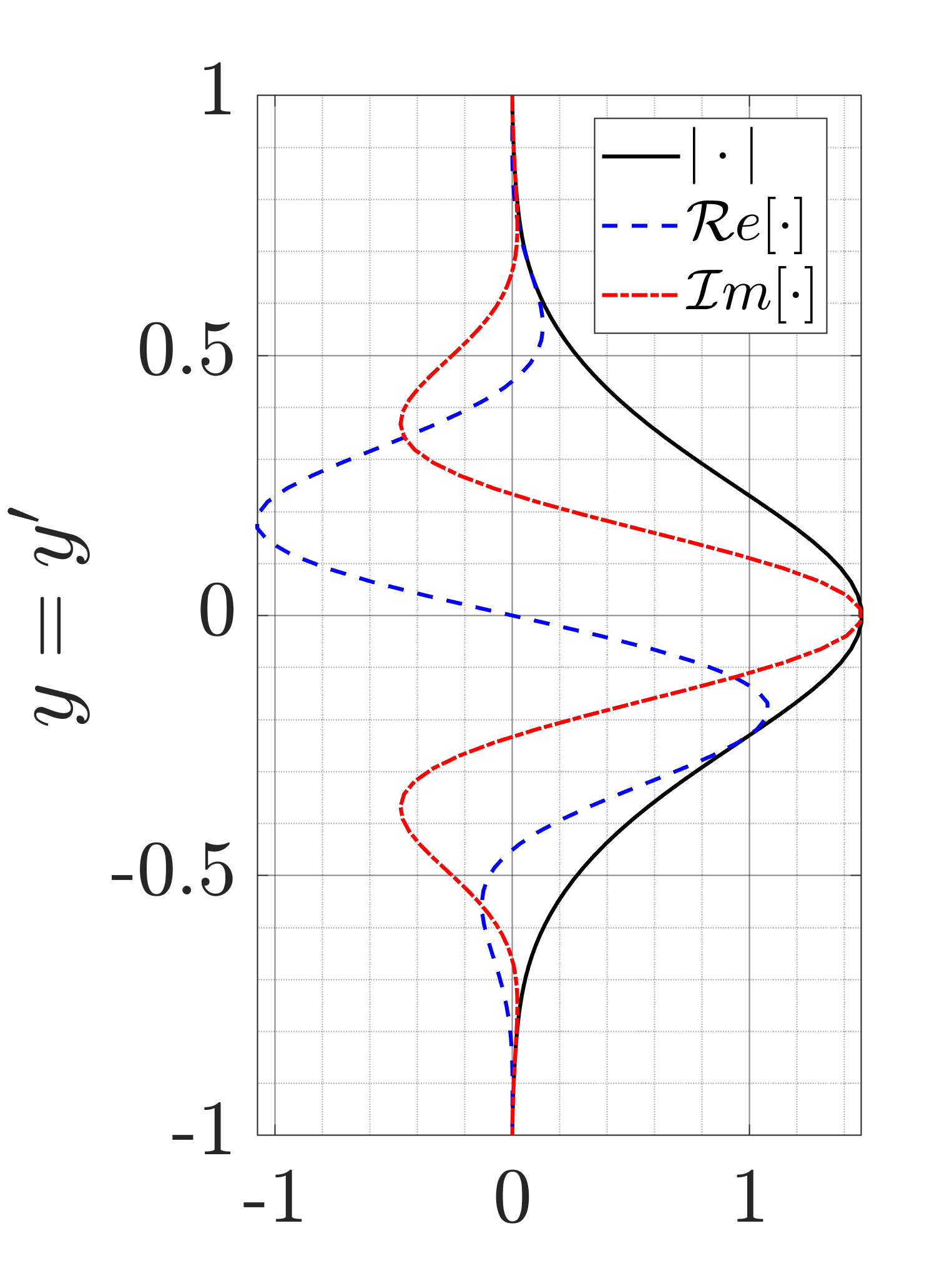}
    \includegraphics[width=0.135\textwidth]{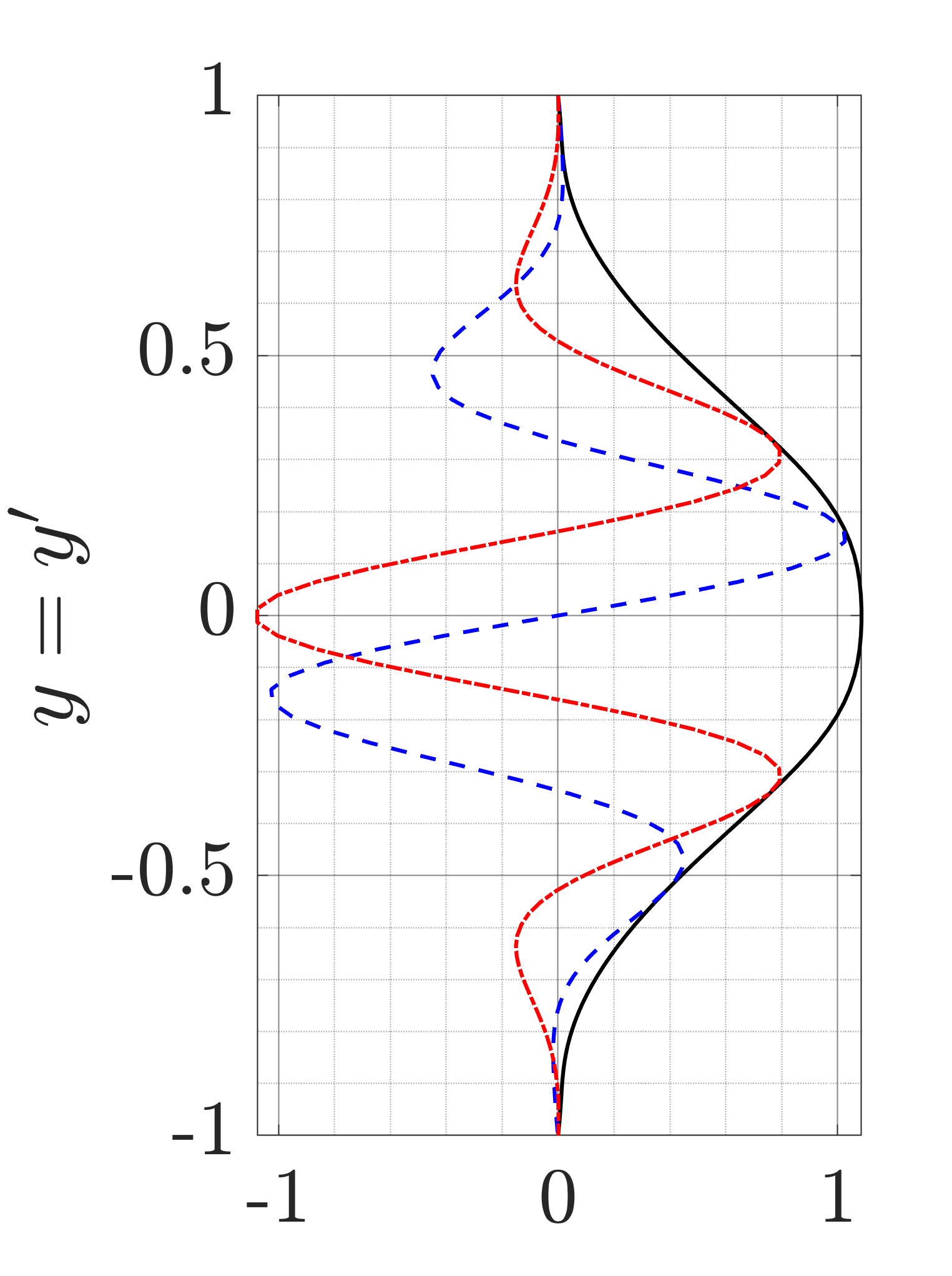}
    \includegraphics[width=0.135\textwidth]{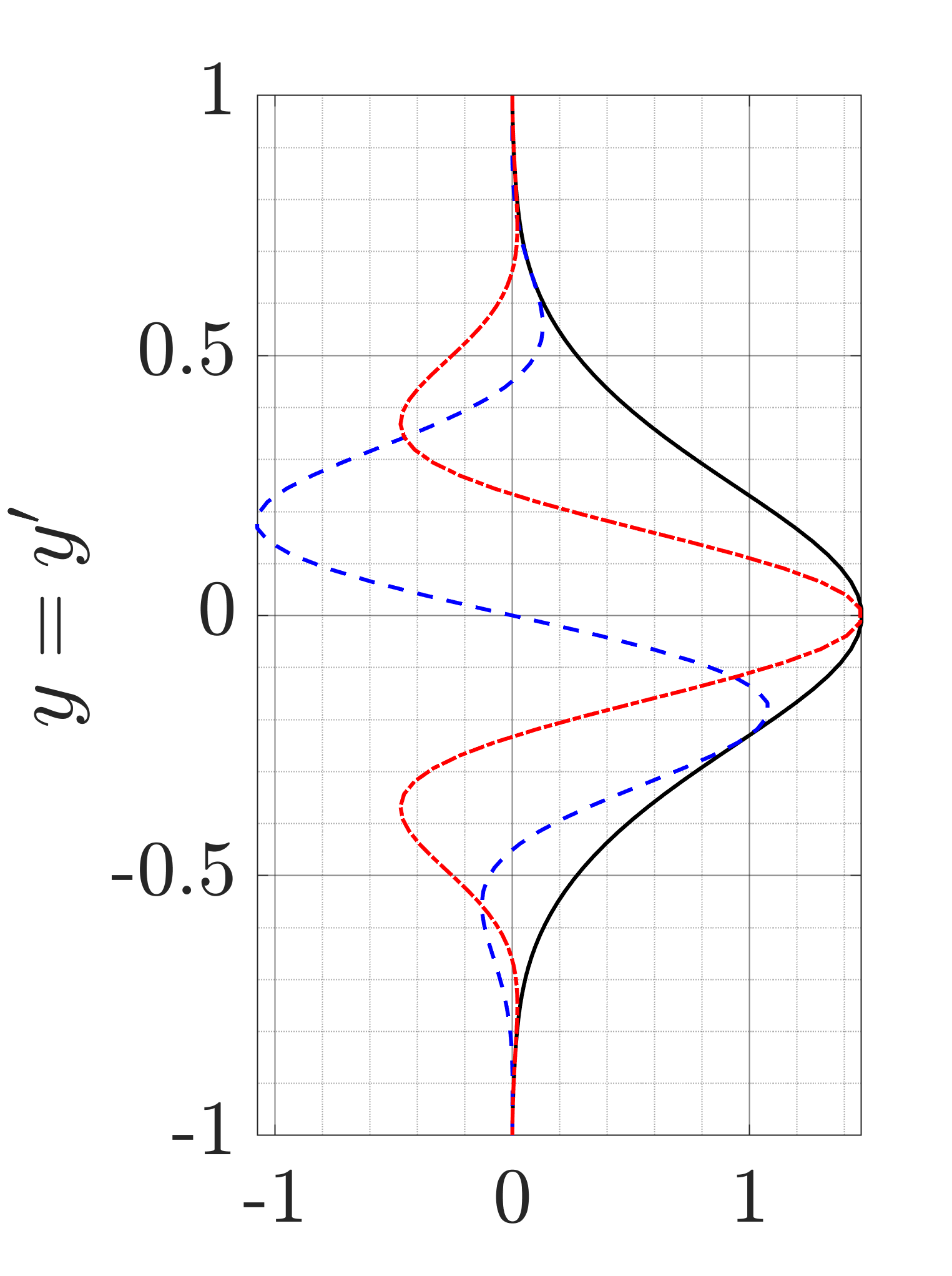}

    \caption{First three components of $\widetilde{\mathbfsbilow{u}}_{\Xi,c}$; i.e.,  (a) $10^3\,\widehat{\mathbfsbilow{u}}_{\xi,1}$, (b) $10^3\,\widehat{\mathbfsbilow{v}}_{\xi,1}$, and (c) $10^3\,\widehat{\mathbfsbilow{w}}_{\xi,1}$ for plane Couette flow at $Re=358$, $k_x=0.22$, $k_z=0.67$ and $\omega=0$.}
    \label{fig:u_xi_diag_2_3_cou}
\end{figure}

Fig. \ref{fig:u_xi_abs_real_imag_diag_1_cou} shows the real and imaginary parts of the streamwise velocity correlations $\widehat{\mathbfsbilow{u}}_{\xi,1}$ for plane Couette flow at $Re=358$ for the wavenumber triplet $k_x=0.22$, $k_z=0.67$ and $\omega=0$ approximately corresponding to the largest $\|\widetilde{\mathcal{H}}_{\nabla}\|_{\mu}(k_x,k_z)$, i.e., the darkest region in Fig. \ref{fig:mu_PCF_358}(a).   Here, we can see that the magnitude of $\widehat{\mathbfsbilow{u}}_{\xi,1}$ shows a peak near the channel center $y\approx 0$. This behavior is consistent with results from secondary instability analysis of streamwise streaks, which show that the least-stable mode is located at the center of the channel \cite[Fig. 8]{reddy1998stability}. Moreover, the real part, shown in Fig. \ref{fig:u_xi_abs_real_imag_diag_1_cou}(a) changes its sign at the channel center, which is also consistent with results from NLOP analysis showing streamwise velocities and streamwise vorticity reversing sign near the channel center \cite[Fig. 7]{cherubini2013nonlinear}. Fig. \ref{fig:u_xi_diag_2_3_cou} shows the autocorrelation (i.e., the correlation at the same wall-normal location $y'=y$) of $\widehat{\mathbfsbilow{u}}_{\xi,1}$, $\widehat{\mathbfsbilow{v}}_{\xi,1}$, and $\widehat{\mathbfsbilow{w}}_{\xi,1}$. Here, we again find that these components show a peak absolute value near the channel center and the real part reversing the sign at the channel center, consistent with observations from NLOP analysis \cite{cherubini2013nonlinear}. Similar trends are observed in the autocorrelations of the other components, with the greatest similarities observed in  $\widehat{\mathbfsbilow{u}}_{\xi,3}$, $\widehat{\mathbfsbilow{v}}_{\xi,3}$ and $\widehat{\mathbfsbilow{w}}_{\xi,3}$, which is consistent with those being associated with a similar peak response in Fig. \ref{fig:three_components_123_cou}. 

\begin{figure}
    \centering
    
    (a) $\mathcal{R}e[10^3\,\widehat{\mathbfsbilow{u}}_{\xi,1}(y,y')]$ \hspace{0.09\textwidth} (b) $\mathcal{I}m[10^3\,\widehat{\mathbfsbilow{u}}_{\xi,1}(y,y')]$
    
    \includegraphics[width=0.23\textwidth]{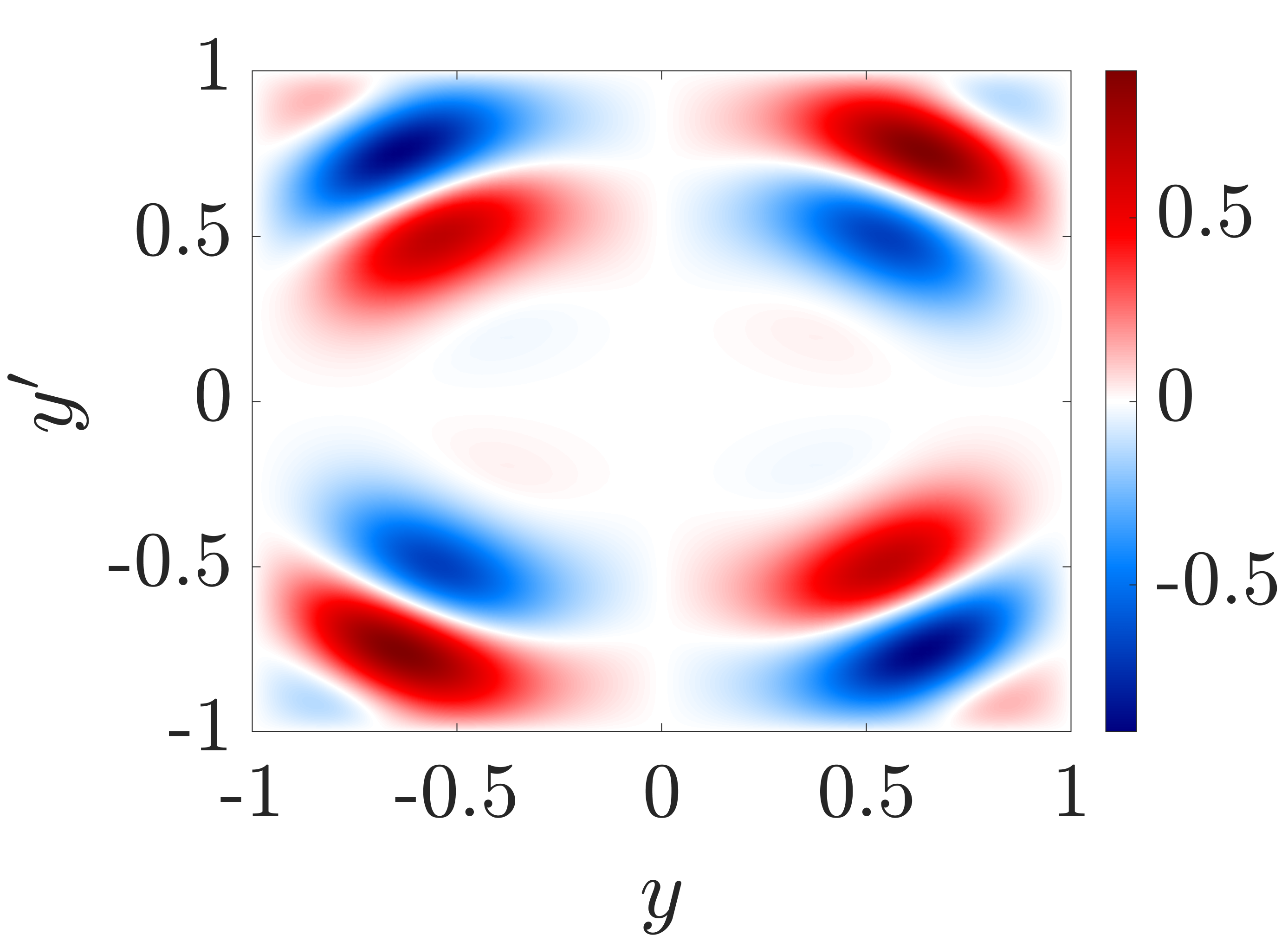}
    \includegraphics[width=0.23\textwidth]{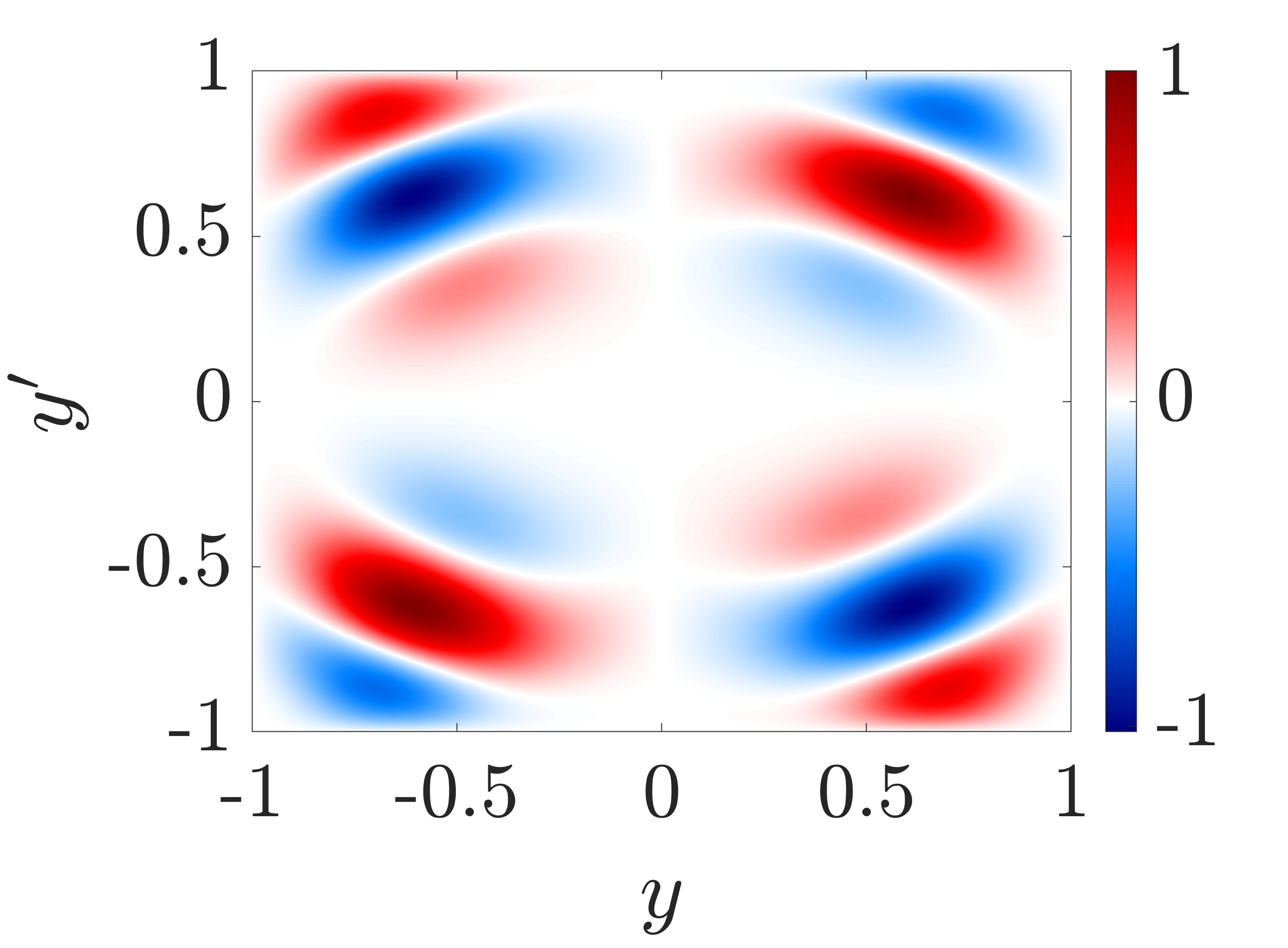}    \caption{$10^3\,\widehat{\mathbfsbilow{u}}_{\xi,1}$ for plane Poiseuille flow at $Re=690$, $k_x=0.65$, $k_z=1.56$ and $c=-\omega/k_x=0.53$.}
    \label{fig:u_xi_abs_real_imag_diag_1_poi}
\end{figure}

\begin{figure}
    \centering
    (a) $10^3\,\widehat{\mathbfsbilow{u}}_{\xi,1}$\hspace{0.05\textwidth} (b) $10^3\,\widehat{\mathbfsbilow{v}}_{\xi,1}$ \hspace{0.05\textwidth} (c) $10^3\,\widehat{\mathbfsbilow{w}}_{\xi,1}$
    
    \includegraphics[width=0.135\textwidth]{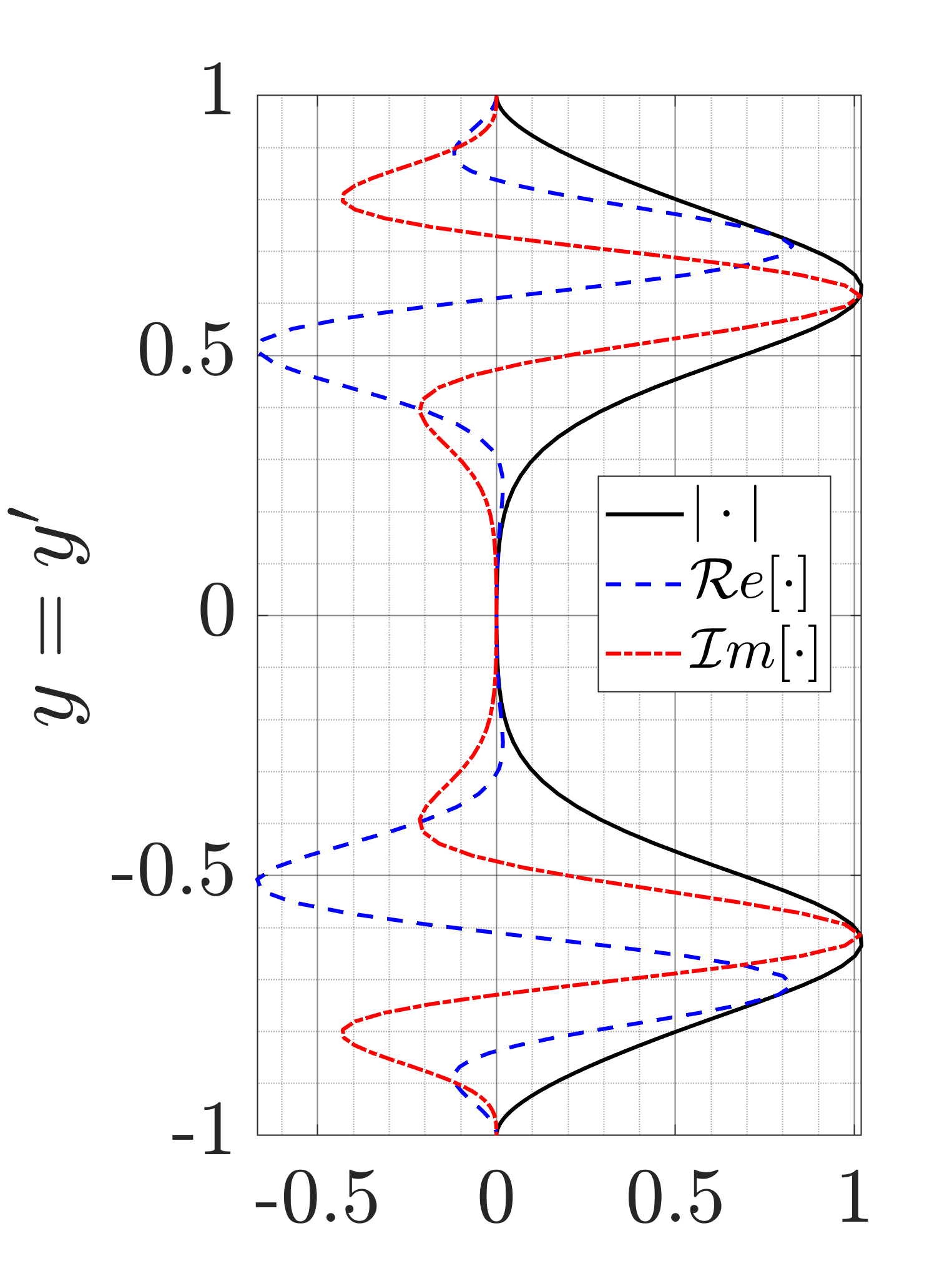}
    \includegraphics[width=0.135\textwidth]{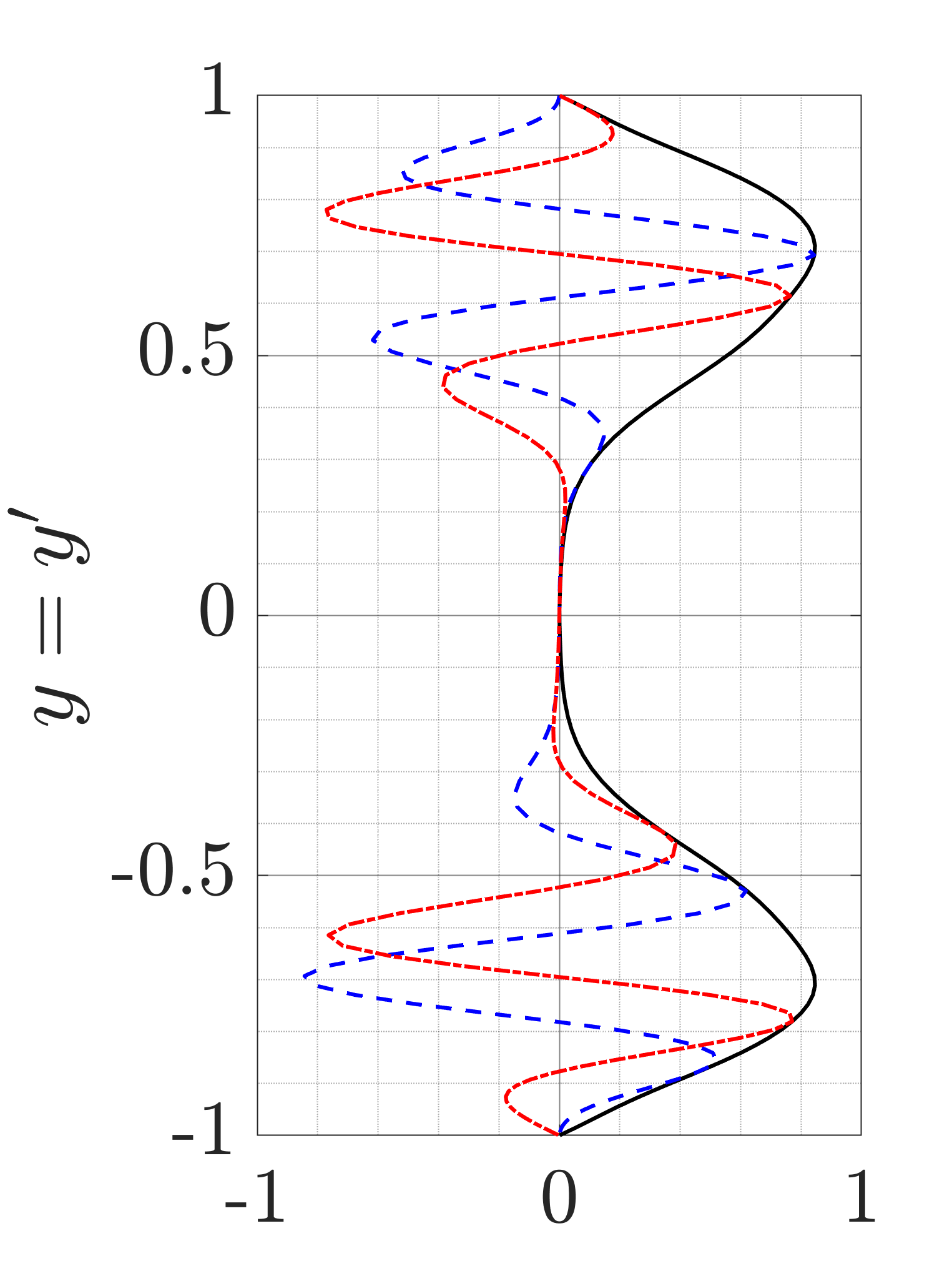}
    \includegraphics[width=0.135\textwidth]{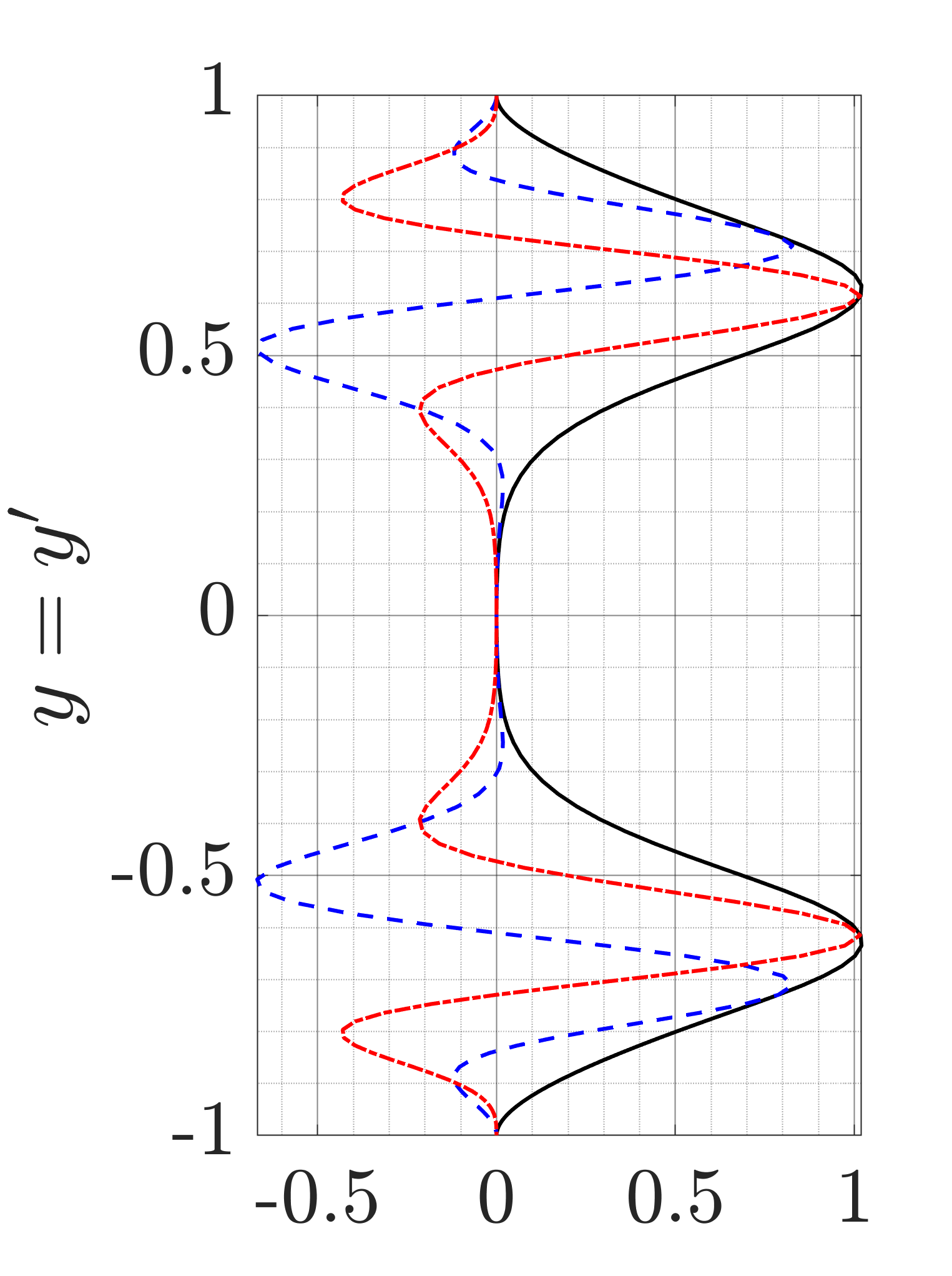}
    
    \caption{First three components of $\widetilde{\mathbfsbilow{u}}_{\Xi,c}$; i.e.,  (a) $10^3\,\widehat{\mathbfsbilow{u}}_{\xi,1}$, (b) $10^3\,\widehat{\mathbfsbilow{v}}_{\xi,1}$, and (c) $10^3\,\widehat{\mathbfsbilow{w}}_{\xi,1}$ for plane Poiseuille flow at $Re=690$, $k_x=0.65$, $k_z=1.56$ and $c=-\omega/k_x=0.53$. 
    }
    \label{fig:u_xi_diag_2_3_poi}
\end{figure}

Fig. \ref{fig:u_xi_abs_real_imag_diag_1_poi} presents $\widehat{\mathbfsbilow{u}}_{\xi,1}$  for plane Poiseuille flow at $Re=690$  corresponding to the wavenumber pair $k_x=0.65$, $k_z=1.56$ showing the largest $\|\widetilde{\mathcal{H}}_{\nabla}\|_{\mu}(k_x,k_z)$ in Fig. \ref{fig:mu_PCF_358}(b). The results are plotted for $c=-\omega/k_x=0.53$, which is the phase speed leading to the largest $\mu_{\mathbfsbilow{\widetilde{U}}_{\Xi,c}}\left[\mathbfsbilow{\widetilde{H}}_{\nabla}(k_x,k_z,\omega)\right]$ at wavenumber pair $k_x=0.65$, $k_z=1.56$. Fig. \ref{fig:u_xi_abs_real_imag_diag_1_poi} shows the real and imaginary parts of $\widehat{\mathbfsbilow{u}}_{\xi,1}$, which are shown to vanish at the channel center ($y=0$). This trend is consistent with observations from NLOP analysis of plane Poiseuille flow \cite[Fig. 3]{parente2022linear} and analysis of optimal secondary energy growth \cite{cossu2007optimal}. Fig. \ref{fig:u_xi_diag_2_3_poi} shows the corresponding autocorrelation $(y'=y)$, where all of the components illustrate a vanishing velocity at the channel center. In contrast, linear optimal perturbations of plane Poiseuille flow peak near the channel center; see e.g. the comparison in \cite{parente2022linear}. Similar behavior is also observed in the other six components, and consistent with our conjecture based on Fig. \ref{fig:three_components_123_poi} that the autocorrelatins and structures associated with $\widehat{\mathbfsbilow{u}}_{\xi,3}$, $\widehat{\mathbfsbilow{v}}_{\xi,3}$ and $\widehat{\mathbfsbilow{w}}_{\xi,3}$  show the greatest overall similarities.  This difference between $\widetilde{\mathbfsbilow{u}}_{\Xi,c}$ in plane Couette flow and plane Poiseuille flow is likely because the laminar base flow $U(y)=y$ in plane Couette flow is odd symmetric over wall-normal locations $y\in [-1,1]$, while $U(y)=1-y^2$ in plane Poiseuille flow is even symmetric.

\section{Conclusions and future work}
\label{sec:conclusion}

This work builds upon recently introduced structured input-output analysis (SIOA) \cite{liu2021structuredJournal} to further examine transitional wall-bounded shear flows. Here we modify the SIOA feedback interconnection to decompose the structured uncertainty operator into block diagonal elements associated with the individual velocity components. We interpret the resulting full block structures as two-point spatial velocity correlations (in the wall-normal direction) associated with the optimal perturbations; i.e. most destabilizing to the imposed feedback interconnection structure. In plane Couette flow the magnitudes of these correlations show maximum values near the channel center in accordance with results from secondary instability analysis of streamwise streaks \cite{reddy1998stability}. The real parts of these correlations reverse the sign near the channel center in a manner similar to NLOP \cite{cherubini2013nonlinear}. For plane Poiseuille flow, the  optimal perturbations show vanishing values near the channel center consistent with NLOP \cite{parente2022linear} and results obtained from optimal secondary energy growth \cite{cossu2007optimal}. In contrast, linear optimal perturbations of plane Poiseuille flow peak near the channel center; see e.g. the comparison in \cite{parente2022linear}. The results provide further evidence that behavior associated with nonlinear effects can be captured through a SIOA framework. 

Applications of this framework to wider parameter regimes and flow configurations such as compressible flows \cite{bhattacharjee2023structured} are directions of ongoing work. Extracting corresponding optimal forcing and response mode shapes \cite{mushtaq2023structured} are expected to provide insights for flow control applications. Another important direction of future study is to exploit the recent work of \cite{mushtaq2022structured}  to enforce equality of the velocity correlations.

\section*{Acknowledgement}
The authors gratefully acknowledge partial support from the US National Science Foundation (CBET 1652244). C.L. would like acknowledge the travel support from Berkeley Postdoctoral Association (BPA) Professional Development Award. 

\raggedend

\bibliography{main}{}
\bibliographystyle{IEEEtran}

\end{document}